\documentstyle[prb,aps,multicol,epsf]{revtex}

\begin{document}

\draft
\title{Spin correlations in the two-leg antiferromagnetic ladder
in a magnetic field}
\author{T. Hikihara}
\address{Department of Earth and Space Science,
Graduate School of Science, Osaka University, \\
Toyonaka, Osaka 560-0043, Japan}
\author{A. Furusaki}
\address{Yukawa Institute for Theoretical Physics, Kyoto University,
Kyoto 606-8502, Japan}
\date{November 20, 2000}
\maketitle
\begin{abstract}
We study the ground-state spin correlations
in the gapless incommensurate regime of a $S=1/2$ $XXZ$ chain
and a two-leg antiferromagnetic ladder under a magnetic field,
in which the gapless excitations form a Tomonaga-Luttinger (TL) liquid.
We calculate numerically the two-spin correlation functions and
the local magnetization in the two models
using the density-matrix renormalization-group method.
By fitting the numerical results for an open $XXZ$ chain of 100
spins to correlation functions of a Gaussian model,
we determine the TL-liquid parameter $K$
and the amplitudes of the correlation functions.
The value of $K$ estimated from the fits is in excellent agreement
with the exact value obtained from the Bethe ansatz.
We apply the same method to the open ladder consisting of 200 spins
and determine the dependence of $K$ on the magnetization $M$.
The $K$-$M$ relation changes drastically
depending on the ratio of the coupling constants
in the leg and rung directions.
We also discuss implications of these results to experiments on
the nuclear spin relaxation rate $1/T_1$ and dynamical spin structure
factors.

\end{abstract}

\begin{multicols}{2}

\section{INTRODUCTION}
Spin ladder systems have been studied extensively over the past
decade.\cite{review}
There are reasons why the ladders have attracted so much attention.
Firstly, they naturally interpolate
one- and two-dimensional systems and may provide some hints
to better understand the high-temperature superconductivity
which occurs in square lattice ${\rm CuO_2}$ planes.
Secondly, spin ladders themselves have interesting physics and
deserve through investigation in their own right.
One of their most surprising properties is that low-energy physics
of spin ladders depends drastically on the number of legs.
Spin-$1/2$ antiferromagnetic (AF) ladders, for example,
have a finite gap in the spin excitation spectrum in the
even-leg case, whereas they have no gap in the odd-leg case.
The ground state of an even-leg ladder is a spin singlet and
its properties can be understood from the
short-range resonating-valence-bond picture.\cite{RVB}
This spin-gap behavior has been observed experimentally
on $S=1/2$ two-leg ladder compounds,\cite{exp1,exp2}
such as ${\rm SrCu_2O_3}$ and ${\rm Cu_2(C_5H_{12}N_2)_2Cl_4}$.

A gapless phase can appear in even-leg ladders when an
external field $h$ is applied.
If the field $h$ is larger than a critical field $h_{c1}$,
which is equal to the spin gap,
and if it is smaller than the saturation field $h_{c2}$,
then the ground state has a nonzero magnetization $M$ and the energy gap
between the ground state and the first-excited states vanishes.
The gapless mode has been shown to be described as a
Tomonaga-Luttinger (TL) liquid both in the strong- and weak-coupling
limit,\cite{Chi-Gia,Gia-Tsv,Furu-Zhn}
where the coupling in the rung-direction $J_\perp$ is much larger or
much smaller than the one in the leg-direction $J_\parallel$,
respectively.
However, the $M$ dependence of the TL-liquid parameter $K$,
which governs the spin correlations in long wave length,
has been obtained analytically only in the strong-coupling limit
and it remains as a nontrivial problem to determine $K$ for general
$J_\perp/J_\parallel$.
In the gapless phase the system shows incommensurate spin correlations
since the Fermi wavenumber of Jordan-Wigner fermions is shifted
from $\pi/2$ in the presence of the magnetic field
which acts as a chemical potential for the fermions.
The wavenumber $Q$ characterizing the incommensurability of the
gapless mode varies continuously as $h$ increases.
This gapless incommensurate (IC) phase is in fact in the same
universality class as the one-dimensional $S = 1/2$ $XXZ$ model
in a magnetic field, as we will see.

In this paper, we study low-energy properties of
the $S = 1/2$ two-leg AF ladder in the gapless IC regime
for broad range of $J_\perp / J_\parallel$.
We show that the system is a TL liquid for arbitrary
$J_\perp / J_\parallel$
and determine the $M$ dependence of $K$ numerically.
To this end, we compute numerically the ground-state spin-correlation
functions and the local magnetization in the open ladders
using the density-matrix renormalization-group (DMRG)
method~\cite{White1,White2}
and extract the TL-liquid parameter by fitting the data to
correlation functions obtained from the Abelian bosonization.
This method was applied in our previous work~\cite{CorAm}
to the $S=1/2$ $XXZ$ chain at $h = 0$ and proved to be effective
in determining both the TL-liquid parameter and amplitudes of
correlation functions.
In order to demonstrate the validity of the analysis in the gapless IC
phase, we first apply it to the $S = 1/2$ $XXZ$ chain for $h > 0$.
The model is exactly solvable by the Bethe ansatz
and the TL-liquid parameter $K$ can be calculated
for arbitrary value of $M$.
It thus provides a good test ground to check
accuracy of our method.
We find that $K$ estimated from the DMRG data is in excellent
agreement with the exact calculation.
We then apply the same method to the two-leg ladders in a magnetic
field.
Our numerical data of correlation functions are fitted well for
broad range of $J_\perp/J_\parallel$ to the formulas based on the
bosonization approach, confirming that the gapless modes are in fact
in the universality class of a TL liquid for arbitrary
$J_\perp/J_\parallel$.
The $M$ dependence of $K$ obtained in the large $J_\perp/J_\parallel$
limit agrees with the analytic result obtained through mapping to the
$XXZ$ chain.
In this limit $K$ is less than 1 for $0 < M < 1$.
As $J_\perp/J_\parallel$ decreases,
$K$ increases and become larger than 1 for intermediate values of $M$.
Our numerical result indicates that $K$ takes a universal value 1 for any
$J_\perp/J_\parallel$ in the limits $M\to0$ and $M\to1$.

The plan of the paper is as follows.
We first review the Abelian bosonization approach
to the $S=1/2$ $XXZ$ chain under a magnetic field in Sec.~II A.
The formulas of the spin correlations and the local magnetization
in finite open chains are presented.
In Sec.~II B, we show numerical data for the $XXZ$ chain of $L=100$
sites obtained from the DMRG calculation and fit the data to the
functions given in Sec.~II A.
In Sec.\ III A, we briefly review some relevant results of the previous
analytic studies on the two-leg ladders
in the strong- and weak-coupling limits.
The DMRG data and the results of fitting on the open ladders
with $L = 200$ sites are shown in Sec.\ III B.
The $M$ dependence of $K$ for various values of $J_\perp/J_\parallel$
is obtained.
Its implications to NMR and neutron scattering experiments are briefly
mentioned.
Finally, our results are summarized in Sec.\ IV.

\section{$XXZ$ CHAIN}
\subsection{Bosonization approach}
In this section, we consider spin-1/2 $XXZ$ chains with open ends in a
magnetic field $h$.
The Hamiltonian is
\begin{equation}
{\cal H}_{\rm ch} = {\cal H}_0 + {\cal H}_h \label{eq:Hchn}
\end{equation}
with
\begin{eqnarray}
{\cal H}_0 &=&
   J \sum_{l=1}^{L-1} (\bbox{S}_l, \bbox{S}_{l+1} )_\Delta,
    \nonumber \\
{\cal H}_h &=& - h \sum_{l=1}^L S^z_l, \nonumber
\end{eqnarray}
where $\bbox{S}_l$ are $S = 1/2$ spin operators and
$(\bbox{S}_l, \bbox{S}_{l'} )_\Delta =
   S^x_l S^x_{l'} + S^y_l S^y_{l'} + \Delta S^z_l S^z_{l'}$.
We assume the system size $L$ to be even throughout this paper
and treat only the case where $J > 0$ and $0 \le \Delta \le 1$.
We note that the Hamiltonian (\ref{eq:Hchn}) for $-1 < \Delta\le 1$
can be solved exactly by Bethe ansatz
for arbitrary values of $h$.\cite{Bethe0,Bethe1,Bethe2}

We use the standard Abelian bosonization techniques to analyze
spin-spin correlation functions at zero temperature.
We basically follow the scheme presented in
Ref.~\onlinecite{Eggert} and generalize it to the case of open chains
in magnetic fields.
The bosonization formulas in the absence of magnetic fields are
reported in Ref.~\onlinecite{CorAm}.

The low-energy dynamics of $XXZ$ chains is described
by the Gaussian model,\cite{Bethe1}
\begin{equation}
\widetilde{{\cal H}}_{\rm ch} =
   \frac{v}{2} \int_0^{L+1} dx
   \left[ \left( \frac{d\phi}{dx}\right)^2
         + \left( \frac{d\tilde{\phi}}{dx}\right)^2 \right] ,
           \label{eq:HchnBos}
\end{equation}
where $v$ is the spin-wave velocity.
The continuous variable $x$ is identified with the site index $l$
under the assumption that the lattice spacing equals unity.
The bosonic fields $\phi(x)$ and $\tilde{\phi}(x)$ obey
the commutation relation
$[\phi(x), \tilde{\phi}(y)] = -i \Theta(x-y)$,
where $\Theta(x)$ is the step function.
The spin operators in the original Hamiltonian (\ref{eq:Hchn}) are
related to the bosonic fields by the relations,
\begin{eqnarray}
S^z_l &=&
\frac{1}{2 \pi R}\frac{d\phi}{dx}
+ a (-1)^l \sin\left(\frac{\phi(l)}{R}\right),
           \label{eq:Szchn} \\
S^{-}_l &=&
\exp[- i 2\pi R\tilde{\phi}(l)]
\left[ b \sin\left(\frac{\phi(l)}{R}\right) + c(-1)^l \right],
      \label{eq:S-chn}
\end{eqnarray}
with $a$, $b$, and $c$ being real constants.
The parameter $R$ determines the exponents of correlation functions.
We also introduce the TL-liquid parameter $K$
by $K = 1/(4\pi R^2)$;
$K=1$ in the $XY$ case ($\Delta=0$), and
$K=1/2$ in the Heisenberg case ($\Delta=1$) at $h=0$.
  From Eq.~(\ref{eq:S-chn}), $S^x_l$ is written as
\begin{equation}
S^x_l = c(-1)^l \cos[2\pi R\tilde{\phi}(l)]
         - i b \sin[2\pi R\tilde{\phi}(l)]
           \sin\left(\frac{\phi(l)}{R} \right).
     \label{eq:Sxchn}
\end{equation}
The second term with the coefficient $ib$ is Hermitian due to the
commutation relation $[\phi(l),\tilde\phi(l)]=-i/2$.
The open boundary conditions are translated to
the boundary conditions on the bosonic fields at the two
phantom sites $l=0$ and $l=L+1$:\cite{Eggert}
$\phi(0)=0$ and $\phi(L+1)=2\pi RLM_{\rm ch}$, where $M_{\rm ch}$ is
the magnetization per site,
\begin{equation}
M_{\rm ch} = \frac{1}{L} \sum_{l=1}^L S^z_l.
\label{eq:Mch}
\end{equation}
The total magnetization $LM_{\rm ch}$ is an integer for even $L$.
These boundary conditions lead to the mode expansion,
\begin{eqnarray}
\phi(x) &=&
   \frac{x}{L+1} \phi_0
   + \sum_{n=1}^\infty \frac{\sin(q_n x)}{\sqrt{\pi n}}
          \left( a_n + a_n^\dagger \right) , \label{eq:mode1} \\
\tilde{\phi}(x) &=&
   \tilde{\phi}_0
   + i \sum_{n=1}^\infty \frac{\cos(q_n x)}{\sqrt{\pi n}}
         \left( a_n - a_n^\dagger \right) ,   \label{eq:mode2}
\end{eqnarray}
where $q_n = \pi n /(L+1)$, $[\tilde{\phi}_0, \phi_0] = i$,
and $a_m$ are boson annihilation operators obeying
$[a_m, a_n^\dagger] = \delta_{m,n}$.
Note that the commutation relation between $\phi(x)$ and
$\tilde{\phi}(y)$ mentioned above is satisfied.
The lowest energy state $|M_{\rm ch}\rangle$ in the subspace
in which the magnetization per spin is $M_{\rm ch}$ is a vacuum of
$a_n$
\begin{equation}
a_n|M_{\rm ch}\rangle = 0
\end{equation}
and an eigenstate of $\phi_0$
\begin{equation}
\phi_0 |M_{\rm ch}\rangle = 2\pi R L M_{\rm ch}|M_{\rm ch}\rangle.
\label{eq:phi_0}
\end{equation}
We may regard $\tilde\phi_0$ as a coordinate variable along a
fictitious ring of radius $1/2\pi R$ and take
$\phi_0=-id/d\tilde\phi_0$ to be its momentum conjugate.
The state $|M_{\rm ch}\rangle$ is then proportional to
$\exp(i2\pi RLM_{\rm ch}\tilde\phi_0)$.
The bosonization formulas (\ref{eq:Szchn}) and
(\ref{eq:S-chn}) represent only the leading contributions.
In the next order $S^-_l$ has a term of the form
$(-1)^l\exp[-2\pi iR\tilde\phi(l)]\cos[2\phi(l)/R]$.
We will, however, ignore this contribution because it yields only
subleading corrections that disappear quickly for large $|l-l'|$.

Using Eqs.~(\ref{eq:Szchn})--(\ref{eq:phi_0}), one can evaluate the
two-spin correlation functions
$\langle S^\alpha_l S^\alpha_{l'} \rangle$ $(\alpha = x, z)$ and
the local magnetization $\langle S^z_l \rangle$ in open chains,
where $\langle \cdots \rangle$ denotes the expectation value
in the state $|M_{\rm ch}\rangle$.
Brief account of their derivation is given in Appendix.
Here we present the final results:
\end{multicols}
\begin{eqnarray}
\langle S^x_l S^x_{l'} \rangle &\equiv&
X(l,l';q) \nonumber\\
&=&
      \frac{f_{\eta/2}(2l) f_{\eta/2}(2l')}{f_\eta(l-l') f_\eta(l+l')}
     \left[ \frac{c^2}{2} (-1)^{l-l'}
    + \frac{bc}{2} {\rm sgn}(l-l') \left(
      \frac{(-1)^l \cos(q l')}{f_{1/2\eta}(2l')}
    - \frac{(-1)^{l'} \cos(q l)}{f_{1/2\eta}(2l)} \right)
     \right. \nonumber \\
&& \left.\hspace*{3cm}
    - \frac{b^2}{4 f_{1/2\eta}(2l) f_{1/2\eta}(2l')}
        \left( \cos\left[ q (l+l')\right]
                  \frac{f_{1/\eta}(l-l')}{f_{1/\eta}(l+l')}
              + \cos\left[ q (l-l')\right]
                \frac{f_{1/\eta}(l+l')}{f_{1/\eta}(l-l')} \right)
   \right],
                         \label{eq:Cxechn} \\
\langle S^z_l S^z_{l'} \rangle &\equiv&
Z(l,l';q) \nonumber\\
&=&
     \frac{q}{2\pi}
     \left(\frac{q}{2\pi}
            + a \frac{(-1)^l \sin(q l)}{f_{1/2\eta}(2l)}
            + a \frac{(-1)^{l'} \sin(q l')}{f_{1/2\eta}(2l')}
     \right)
    - \frac{1}{4\pi^2 \eta} \left(
            \frac{1}{f_2(l-l')} + \frac{1}{f_2(l+l')} \right)
            \nonumber \\
    && + \frac{a^2}{2}\frac{(-1)^{l+l'}}{f_{1/2\eta}(2l)f_{1/2\eta}(2l')}
          \left( \cos\left[q (l-l')\right]
                       \frac{f_{1/\eta}(l+l')}{f_{1/\eta}(l-l')}
               - \cos\left[q (l+l')\right]
                  \frac{f_{1/\eta}(l-l')}{f_{1/\eta}(l+l')} \right)
             \nonumber \\
    && + \frac{a}{2\pi \eta} \left(
          \frac{(-1)^l \cos(q l)}{f_{1/2\eta}(2l)}
                   [g(l+l')+g(l-l')]
      + \frac{(-1)^{l'} \cos(q l')}{f_{1/2\eta}(2l')}
                   [g(l+l')-g(l-l')]  \right) ,
      \label{eq:Czechn} \\
\langle S^z_l \rangle &\equiv&
z(l;q)=
\frac{q}{2\pi}
   + a \frac{(-1)^l \sin(q l)}{f_{1/2\eta}(2l)},
         \label{eq:Szechn}
\end{eqnarray}
\begin{multicols}{2}
where
\begin{eqnarray}
&&
\eta=2\pi R^2=\frac{1}{2K},\\
&&
f_\alpha(x) =
   \left[
\frac{2(L+1)}{\pi}\sin\left(\frac{\pi |x|}{2(L+1)} \right)
\right]^\alpha,
   \label{eq:fx} \\
&&
g(x) = \frac{\pi}{2(L+1)} \cot\left( \frac{\pi x}{2(L+1)}\right).
          \label{eq:gx}
\end{eqnarray}
The wavenumber $q$, characterizing the IC character of
the spin correlations in a magnetic field, is related to $M_{\rm ch}$
by
\begin{equation}
q = \frac{2\pi M_{\rm ch}L}{L+1}.
\label{eq:Qch}
\end{equation}
The factor $L/(L+1)$ appears as a
result of the open boundary conditions.
Under the periodic boundary conditions $q$ should be simply
equal to $2\pi M_{\rm ch}$, because the first term in
Eq.~(\ref{eq:mode1}) is $\phi_0 x/L$ in this case.
This term must be $\phi_0 x/(L+1)$ in the open-boundary case in order
for $\phi(x)$ and $\tilde\phi(x)$ to satisfy the commutation relation
in the interval $[0,L+1]$.
We emphasize that Eqs.~(\ref{eq:Czechn}) and (\ref{eq:Szechn})
reproduce the exact results for the $XY$ chain when $\eta=1/2$ and
$a=-1/\pi$.
As is well known, the $XXZ$ spin chain is equivalent to a model of
spinless fermions with nearest-neighbor interaction.
In this model, $S^z_l$ is none but the fermion density, and
the oscillating term in Eq.~(\ref{eq:Szechn}) corresponds to the
Friedel oscillations near the open
ends.\cite{Friedel1,Friedel2,Friedel3,Friedel4}

In the thermodynamic limit ($L \to \infty$) with $|l-L/2| \ll L$ and
$|l'-L/2| \ll L$, the spin correlations have the asymptotic forms
\begin{eqnarray}
\langle S^x_l S^x_{l'} \rangle &=&
   A_x\frac{(-1)^{l-l'}}{|l-l'|^\eta}
   - \widehat{A}_x \frac{\cos\left[q (l-l')\right]}
                        {|l-l'|^{\eta+1/\eta}},
            \label{eq:CxLchn} \\
\langle S^z_l S^z_{l'} \rangle &=& {M_{\rm ch}}^2
   + A_z (-1)^{l-l'} \frac{\cos\left[q (l-l')\right]}
                         {|l-l'|^{1/\eta}}  \nonumber \\
&&- \frac{1}{4\pi^2 \eta |l-l'|^2},
            \label{eq:CzLchn}
\end{eqnarray}
where the correlation amplitudes $A_x$, ${\widehat{A}_x}$, and $A_z$
are related to the numerical constants
$a$, $b$, and $c$ by $A_x = c^2/2$,
${\widehat{A}_x} = b^2/4$, and $A_z = a^2/2$.
We can therefore estimate the TL-liquid parameter and
the correlation amplitudes in the thermodynamic limit
by extracting the fitting parameters $R$, $a$, $b$, and $c$
from the numerical data on a finite system with use of
Eqs.~(\ref{eq:Cxechn}), (\ref{eq:Czechn}), and (\ref{eq:Szechn}).
At the same time, the TL-liquid parameter $K$ can be calculated exactly for
any $M_{\rm ch}$ by solving an integral equation obtained from the
Bethe ansatz.\cite{TLpr1,TLpr2}
We will compare our estimates of $K$ obtained from the fitting procedure
with the Bethe ansatz results in the next subsection.

\subsection{Numerical results}
Using the DMRG method,\cite{White1,White2}
we computed the two-spin correlation functions
$\langle S^\alpha_l S^\alpha_{l'} \rangle$ $(\alpha=x,z)$ and
the local magnetization $\langle S^z_l \rangle$ in the $L=100$ open
chains.
The two-point functions were calculated for $l = r_0 - r/2$ and
$l' = r_0 + r/2$, where $r_0 = L/2$ for even $r$ and $r_0 = (L+1)/2$
for odd $r$.
The calculation was performed for each lowest-energy state of
${\cal H}_0$ in the subspace of various values of $M_{\rm ch}$.
We employed the finite system algorithm of improved
version.\cite{White3}
The maximum number of kept states $m$ is $100$.
We estimate the numerical error due to the truncation of the Hilbert
space from the difference between the data with $m = 100$ and those
with $m = 70$.
The estimated errors for $\langle S^x_l S^x_{l'} \rangle$,
$\langle S^z_l S^z_{l'} \rangle$, and $\langle S^z_l \rangle$ are,
at largest, of order $10^{-5}$,$10^{-6}$, and $10^{-6}$, respectively.

\begin{figure}
\epsfxsize=80mm
\epsfbox{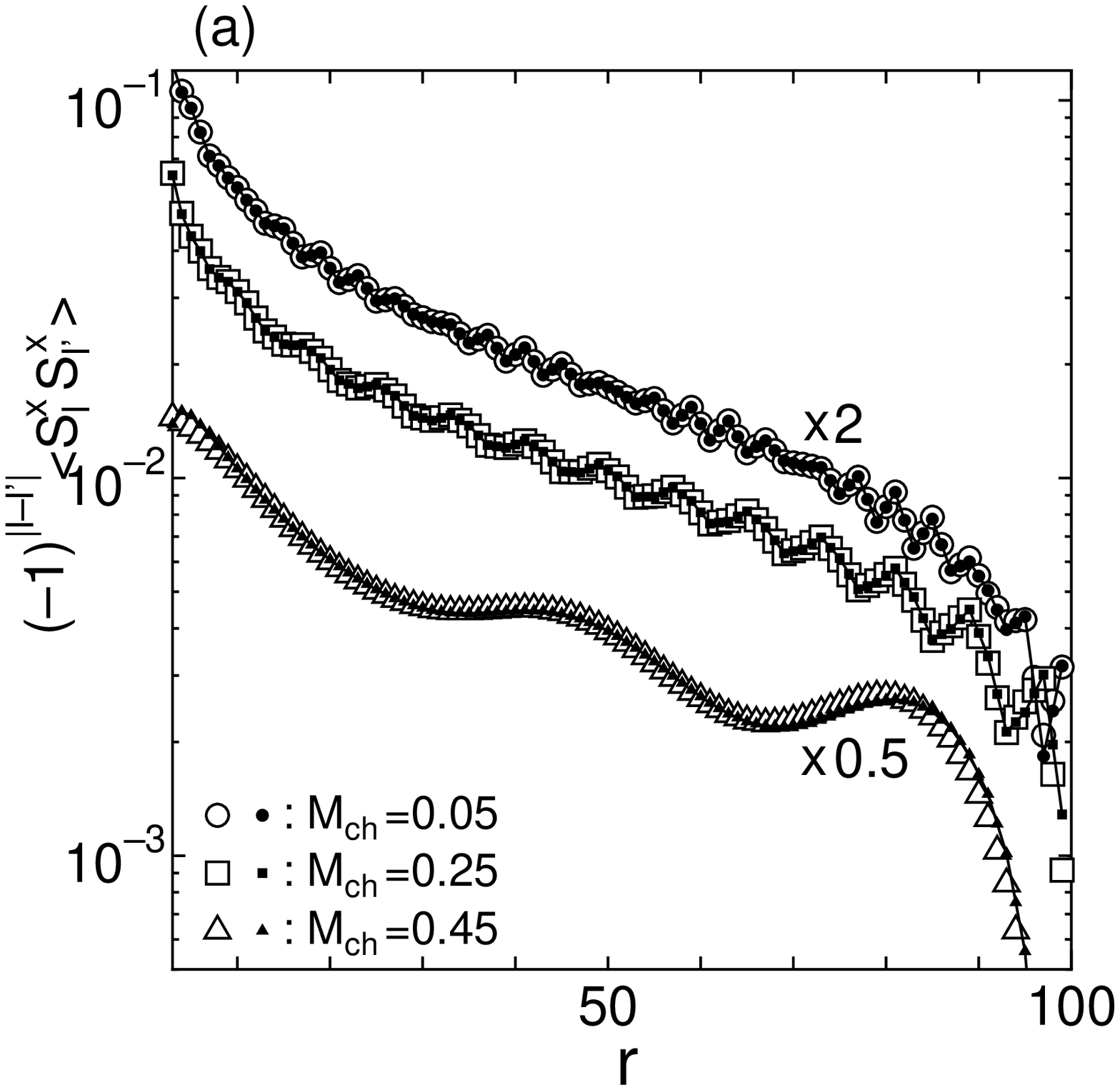}
\epsfxsize=80mm
\epsfbox{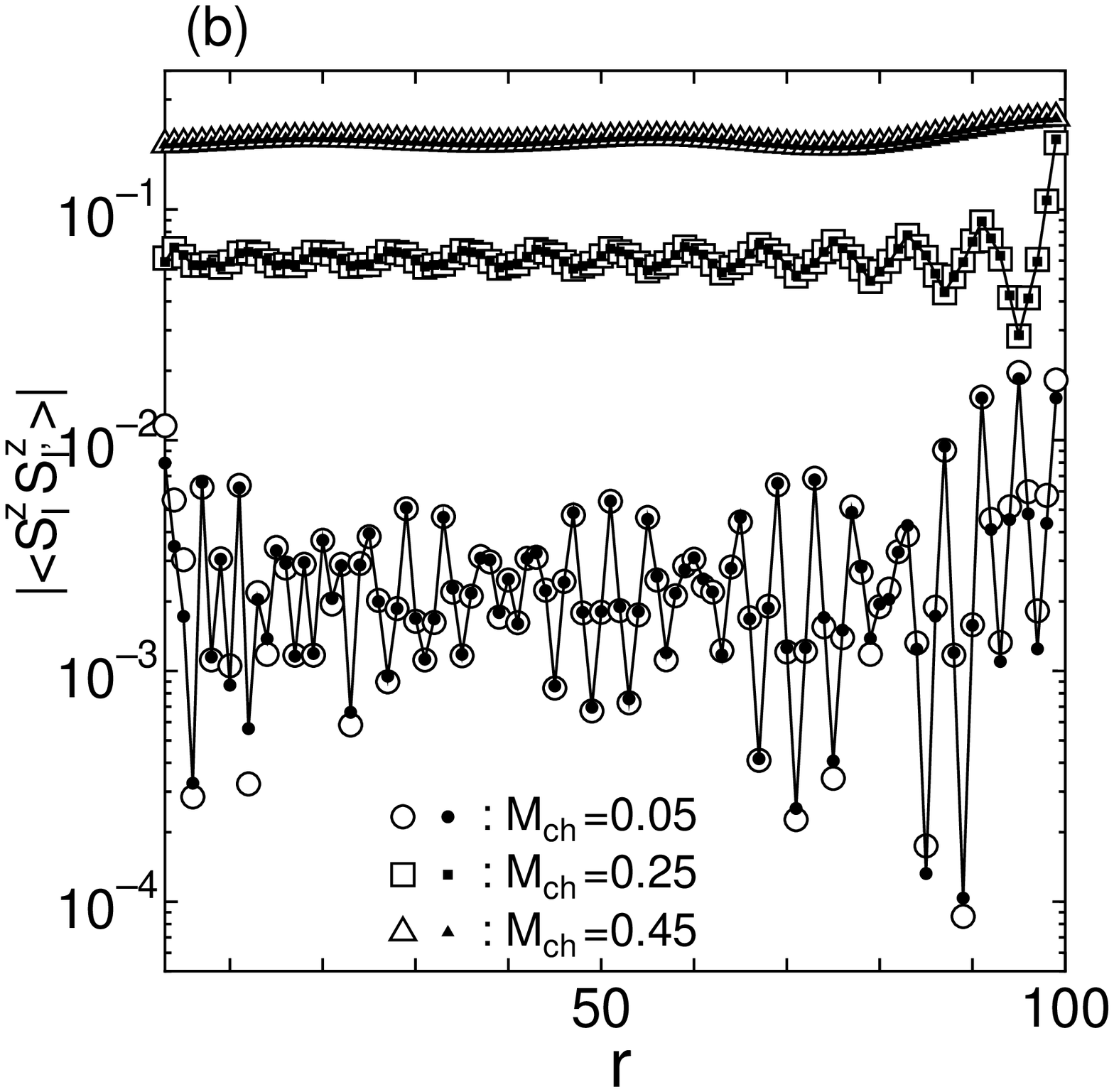}
\epsfxsize=80mm
\epsfbox{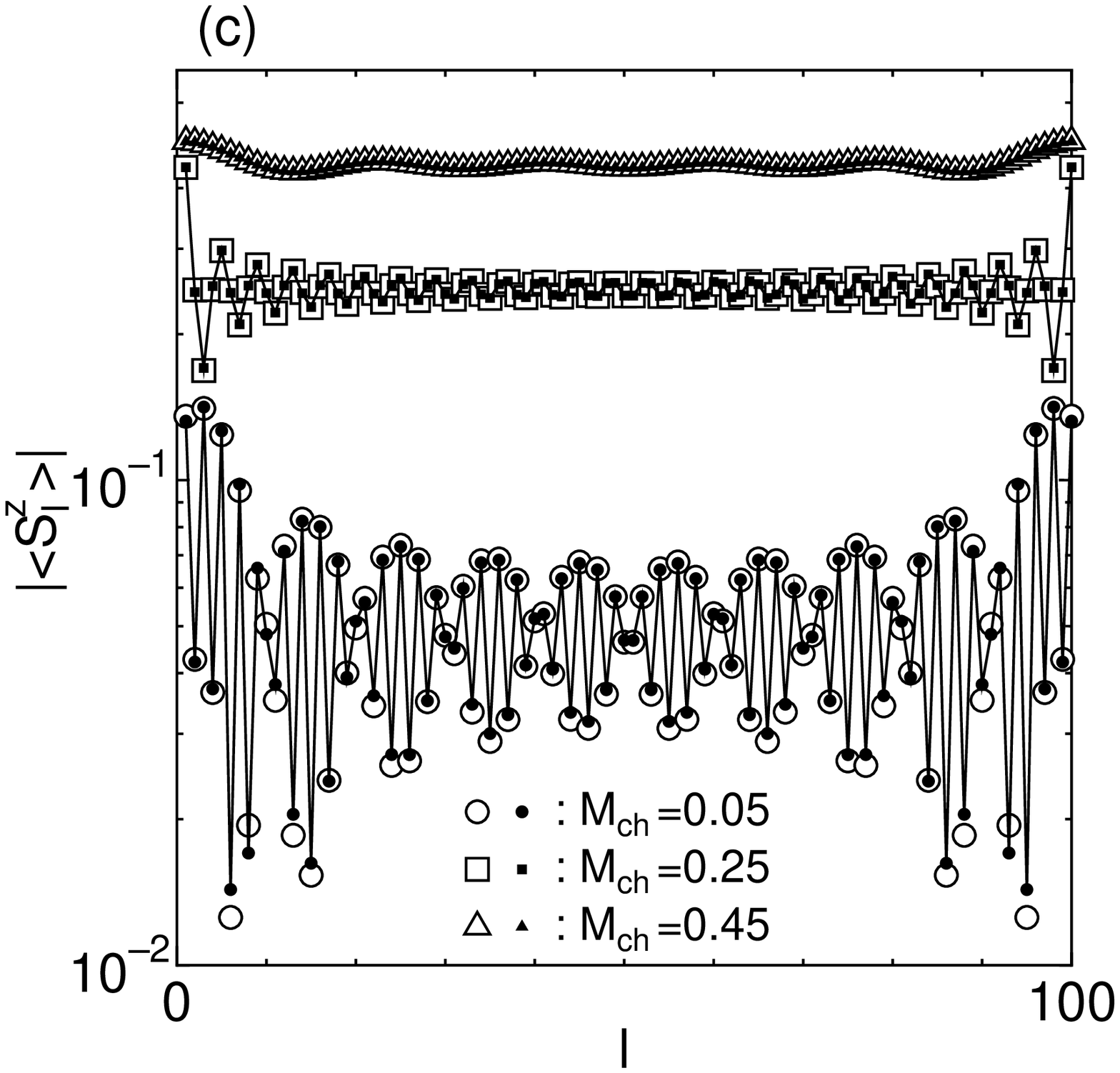}
\narrowtext
\caption{(a) $(-1)^{|l-l'|} \langle S_l^x S_{l'}^x \rangle$
versus $r = |l-l'|$,
(b) $|\langle S_l^z S_{l'}^z \rangle|$ versus $r$,
(c) $\langle S_l^z \rangle$ versus $l$ for $\Delta = 0.5$
and $M_{\rm ch}=0.05$, 0.25, and 0.45.
The open symbols are the DMRG data and small solid symbols are
the fits.
The numerical errors of the DMRG data are smaller than the size of
the open symbols.
The data of $(-1)^{|l-l'|} \langle S_l^x S_{l'}^x \rangle$
for $M_{\rm ch}=0.05$ and $0.45$ in figure (a)
are multiplied by a factor of $2$ and $0.5$, respectively.}
\label{fig:chn}
\end{figure}

In Fig.~\ref{fig:chn}, we show the spin correlations
$\langle S_l^\alpha S_{l'}^\alpha \rangle$
($\alpha = x,z$) and the local magnetization $\langle S_l^z \rangle$
at $\Delta = 0.5$ for three different values of $M_{\rm ch}$.
The DMRG data are shown by open symbols whose sizes
are larger than the truncation error mentioned above.
Taking $R$, $a$, $b$, and $c$ as fitting parameters, we fit
the numerical data to Eqs.~(\ref{eq:Cxechn})--(\ref{eq:Szechn}).
The results of the fitting using the DMRG data of
$\langle S_l^\alpha S_{l'}^\alpha \rangle$
for $10 \le r \le 90$ and of $\langle S_l^z \rangle$ for
$1 \le l \le 100$ are also plotted in the figure by the small solid
symbols.
One can see that the fits are in excellent agreement with
the DMRG data, proving the validity of the bosonization formulas
(\ref{eq:Cxechn})--(\ref{eq:Szechn}).

\begin{figure}
\epsfxsize=80mm
\epsfbox{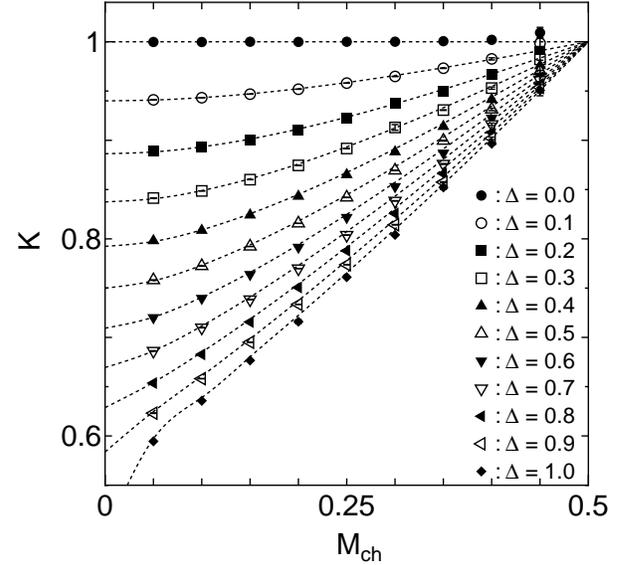}
\caption{The results of $K = 1/(4 \pi R^2)$ estimated from the fitting
of $\langle S^x_l S^x_{l'} \rangle$ for $0 \le \Delta \le 1.0$.
The dotted curves represent the exact values obtained from Bethe ansatz.}
\label{fig:Kchn}
\end{figure}

For various values of $M_{\rm ch}$ and $\Delta$, we determined the
parameters $R$, $a$, $b$, and $c$ by the fitting procedure.
In doing so we used numerical data of several ranges,
$10 \le r \le 80$, $10 \le r \le 90$,
$20 \le r \le 80$, and $20 \le r \le 90$ for the two-spin correlation
functions and $1 \le l \le 100$ and $10 \le l \le 90$ for
the local magnetization.
We take the mean and the variance of the fitting parameters obtained
for the different ranges of $r$ and $l$
as the estimated value and the error of the estimates,
respectively.
The TL-liquid parameter $K \equiv 1/(4\pi R^2)$ estimated from the
numerical data of $\langle S^x_l S^x_{l'} \rangle$
is plotted as a function of $M_{\rm ch}$ in Fig.~\ref{fig:Kchn}.
The exact values obtained from the Bethe ansatz
method~\cite{TLpr1,TLpr2} are also shown as dotted lines.
The agreement is excellent.
We also estimated $K$ from the fitting of $\langle S^z_l S^z_{l'}\rangle$
and $\langle S^z_l\rangle$ and obtained similar results as
Fig.~\ref{fig:Kchn}.
We found, however, that the estimates from the last two correlators
show some deviations from the Bethe ansatz results when $K$
is small.
We do not exactly know why they deviate.
One possible reason might be the effect of the leading irrelevant
operator neglected in the
Gaussian model that becomes marginal at $K=1/2$.
In the $XY$ regime of our interest,
spins have stronger correlations in the $S^x$ and $S^y$ components
than in $S^z$, and thus we may expect that
$\langle S^x_l S^x_{l'}\rangle$ should give us most reliable estimates.

For the correlation amplitudes $A_x$ and $A_z$,
Lukyanov and Zamolodchikov conjectured the exact formulas which are
valid at $h=0$,\cite{Lu-Za,Luky}
\end{multicols}
\begin{eqnarray}
A_x^{\rm LZ} &=&
\frac{1}{8(1-\eta)^2}
\left[\frac{\Gamma(\frac{\eta}{2(1-\eta)})}
             {2\sqrt{\pi}\,\Gamma(\frac{1}{2(1-\eta)})}
\right]^\eta
\exp\left[
-\int^\infty_0\frac{dt}{t}
   \left(\frac{\sinh(\eta t)}{\sinh(t)\cosh[(1-\eta)t]}
         -\eta e^{-2t}\right)\right],
\label{eq:LZx} \\
A_z^{\rm L} &=&
\frac{2}{\pi^2}
\left[\frac{\Gamma(\frac{\eta}{2(1-\eta)})}
             {2\sqrt{\pi}\,\Gamma(\frac{1}{2(1-\eta)})}
\right]^{1/\eta}
\exp\left[
   \int^\infty_0\frac{dt}{t}
   \left(\frac{\sinh[(2\eta-1) t]}{\sinh(\eta t)\cosh[(1-\eta)t]}
         -\frac{2\eta -1}{\eta} e^{-2t}\right)\right],
\label{eq:LZz}
\end{eqnarray}
\begin{multicols}{2}\noindent
where $\Gamma(x)$ is the Gamma function.
Equations (\ref{eq:LZx}) and (\ref{eq:LZz}) have been confirmed
numerically.\cite{CorAm,Luky}
In Table~\ref{tab:Achn}, we give our estimates of
the correlation amplitudes
$A_x = c^2/2$ and $A_z = a^2/2$ obtained from
the fitting of $\langle S^x_l S^x_{l'}\rangle$
and $\langle S^z_l \rangle$ for $0 < M_{\rm ch} < 0.5$,
together with the exact values (\ref{eq:LZx}) and (\ref{eq:LZz}) at
$M_{\rm ch} = 0$.~\cite{no hat Ax}
As can be seen in Table~\ref{tab:Achn} (a),
$A_x$ decreases monotonically from the value given by
Eq.~(\ref{eq:LZx}) to zero as $M_{\rm ch}$ increases from 0 to $1/2$.
Thus, $A_x$ depends not only on $K$ but also on $M_{\rm ch}$.
(See, for example, the data for $\Delta=0$ where $K$
takes a constant value 1 for any $M_{\rm ch}$.)
When $M_{\rm ch}$ approaches $1/2$, where $K\to1$, $A_x$ seems to go
to zero linearly for any $\Delta$.
This can be easily understood once we consider one-magnon contribution
to the correlation function.
On the other hand, $A_z$ decreases monotonically from
the number given by Eq.~(\ref{eq:LZz})
to the universal value $A_z = 1/(2\pi^2) \simeq 0.05066$
as $M_{\rm ch}$ increases from 0 to $1/2$.~\cite{Ampnote}
An exception is the case $\Delta = 0$,
where $A_z = 1/(2\pi^2)$ for any $M_{\rm ch}$.
The convergence of $A_z$ to $1/(2\pi^2)$ at $M_{\rm ch} \to 1/2$
is consistent with the fact that
the correlator $\langle S^z_l S^z_{l'}\rangle$
must take a constant value $1/4$ at $M_{\rm ch}=1/2$.
The right-hand side of Eq. (\ref{eq:CzLchn}) equals
${M_{\rm ch}}^2$ when $\eta = 1/2$, $q = \pi$, and $A_z = 1/(2\pi^2)$.

\section{TWO-LEG AF LADDER}
Encouraged by the success in the last section, we study the two-leg AF
ladders in a magnetic field using the same method.
We begin with a brief review of the analytic results on the ladder
in the strong- and weak-coupling limits.

\subsection{Review of Analytic Results}
The Hamiltonian of the open two-leg ladder studied in this section is
given by
\begin{eqnarray}
{\cal H} & = &
   J_\parallel \sum_{\mu = 1,2} \sum_{l=1}^{L-1}
    (\bbox{S}_{\mu,l},\bbox{S}_{\mu,l+1} )_\Delta
   + J_\perp \sum_{l=1}^L (\bbox{S}_{1,l},\bbox{S}_{2,l} )_\Delta
   \nonumber \\ &&
   - h \sum_{\mu=1,2} \sum_{l=1}^L S_{\mu,l}^z.  \label{eq:Hlad}
\end{eqnarray}
The anisotropy $\Delta$ is introduced for generality.
We assume that the coupling in the leg- and rung-direction,
$J_\parallel$ and $J_\perp$, are positive (antiferromagnetic).
The spin ladder has an excitation gap in weak magnetic fields
$h < h_{c1}$.
We concentrate on the ladder in the gapless IC regime, i.e., in
the case $h_{c1} \le h \le h_{c2}$.
We denote the ratio $J_\perp / J_\parallel$ by $j$ hereafter.

We begin with the strong-coupling limit ($j \gg 1$), for which
a simple intuitive picture is available.
It is known that the system in this limit can be mapped to an
effective $S=1/2$ $XXZ$ chain,\cite{Furu-Zhn,Mila,Totsu} as we
explain below.
Let us first assume $J_\parallel = 0$.
In this case, an eigenstate of ${\cal H}$ is written as
a direct product of rung states.
At each rung two spins $\bbox{S}_{1,l}$
and $\bbox{S}_{2,l}$ are either in a singlet state
$|s_l \rangle = ( |\uparrow \downarrow \rangle
                   - |\downarrow \uparrow \rangle ) / \sqrt{2}$
or in one of the triplet states,
$|t_l^+ \rangle = |\uparrow \uparrow \rangle$,
$|t_l^0 \rangle = ( |\uparrow \downarrow \rangle
                   + |\downarrow \uparrow \rangle ) / \sqrt{2}$,
and $|t_l^- \rangle = |\downarrow \downarrow \rangle$.
When $h$ is small, the ground state consists of a product of
the singlet rungs.
As the field $h$ increases, the energy of the state
$|t_l^+ \rangle$ becomes lower,
and at $h = J_\perp(1+\Delta)/2$, the state degenerates with
$|s_l\rangle$.
We can thus analyze the low-energy properties of the system
for $h \simeq J_\perp(1+\Delta)/2$ by retaining only
the two lowest-energy states $|s_l\rangle$ and $|t_l^+ \rangle$
for each rung.
We may regard the two states as ^^ ^^ down" and ^^ ^^ up" states
of an effective $S = 1/2$ spin,
\begin{eqnarray}
\widetilde{S}^x_l&=&-\frac{1}{\sqrt{2}} (S^x_{1,l}-S^x_{2,l}),
\quad
\widetilde{S}^y_l=-\frac{1}{\sqrt{2}} (S^y_{1,l}-S^y_{2,l}),\\
\widetilde{S}^z_l&=&S^z_{1,l}+S^z_{2,l}-\frac{1}{2}.
\label{tilde S^z}
\end{eqnarray}
The effective spins $\widetilde{\bbox{S}}_l$ are the only low-energy
degrees of freedom, and their dynamics is governed by
\begin{eqnarray}
\widetilde{{\cal H}}
        &=&
   J_\parallel \sum_{l=1}^{L-1}
     (\widetilde{\bbox{S}}_l, \widetilde{\bbox{S}}_{l+1} )_{\Delta/2}
        - \frac{\Delta}{4} J_\parallel
               \left( \widetilde{S}_1^z + \widetilde{S}_L^z \right)
   \nonumber\\ &&
   - \left( h - \frac{1+\Delta}{2} J_\perp
            - \frac{\Delta}{2} J_\parallel
              \right) \sum_{l=1}^L \widetilde{S}_l^z
    + {\rm const}.
        \label{eq:Hefstr}
\end{eqnarray}
This mapping is derived in lowest order in $J_\parallel/J_\perp$ and
valid for the entire IC region of $0<M<1$, where $M$ is magnetization
per rung
\begin{equation}
M=\frac{1}{L}\sum_{l=1}^L\left(S^z_{1,l}+S^z_{2,l}\right).
\label{M}
\end{equation}
Note that the anisotropy of the effective $S=1/2$ $XXZ$ chain
is a half of the anisotropy of the original ladder, $\Delta/2$.
Besides the bulk effective magnetic field
$h - (1+\Delta) J_\perp /2 - \Delta J_\parallel /2$,
there is an additional field $- \Delta J_\parallel/4$
applied only to the boundary spins
$\widetilde{S}_1^z$ and $\widetilde{S}_L^z$.
It induces oscillating magnetization near the boundaries
superposed on the Friedel oscillation which is already present at
any $M\ne0$ without the boundary field.
The effect of the boundary field may be cancelled by adding an extra
term
\begin{equation}
{\cal H}' =
   h' \sum_{\mu = 1,2} \left( S^z_{\mu,1} + S^z_{\mu,L} \right)
\label{eq:H'}
\end{equation}
to the original ladder Hamiltonian (\ref{eq:Hlad}),
where $h' = \Delta J_\parallel/4$ for $j \gg 1$.
Now we define
\begin{eqnarray}
\bbox{S}_{0,l} &=& \bbox{S}_{1,l} + \bbox{S}_{2,l},
   \label{eq:S0} \\
\bbox{S}_{\pi,l} &=& \bbox{S}_{1,l} - \bbox{S}_{2,l}.
   \label{eq:Spi}
\end{eqnarray}
  From the mapping explained above, we conclude that the two-spin
correlations $\langle S_{\pi,l}^x S_{\pi,l'}^x \rangle$
and $\langle S_{0,l}^z S_{0,l'}^z \rangle$
and the local magnetization $\langle S_{0,l}^z\rangle$ in the open
ladder ${\cal H} + {\cal H}'$ in the limit $j\gg1$ are given by the
corresponding correlators in the $XXZ$ chain.
We thus obtain
\begin{eqnarray}
\langle S_{\pi,l}^x S_{\pi,l'}^x \rangle &=&
   2 X(l,l';Q),
        \label{eq:Cxlad} \\
\langle S_{0,l}^z S_{0,l'}^z \rangle &=&
   \left\langle
    \left(\frac{1}{2}+\widetilde{\bbox{S}}_l^z\right)
    \left(\frac{1}{2}+\widetilde{\bbox{S}}_{l'}^z\right)
   \right\rangle
   \nonumber\\
   &=& \frac{1}{4}+\frac{1}{2}[z(l;Q)+z(l';Q)]+Z(l,l';Q),
        \label{eq:Czlad} \\
\langle S_{0,l}^z\rangle &=&
   \frac{1}{2}+z(l,Q),  \label{eq:Szlad}
\end{eqnarray}
where the wavenumber is
\begin{equation}
Q = \frac{2\pi L}{L+1}\left(M-\frac{1}{2}\right).  \label{eq:Qlad}
\end{equation}
In the limit $L\to\infty$ the two-spin correlation functions reduce
to
\begin{eqnarray}
\langle S_{\pi,l}^x S_{\pi,l'}^x\rangle &=&
2A_x\frac{(-1)^{l-l'}}{|l-l'|^{1/2K}} \nonumber\\
&&
-2\widehat{A}_x(-1)^{l-l'}
 \frac{\cos[2\pi M(l-l')]}{|l-l'|^{2K+(1/2K)}},
\\
\langle S_{0,l}^z S_{0,l'}^z\rangle &=&
M^2-\frac{1}{4\pi^2\eta|l-l'|^2} \nonumber\\
&&
+A_z\frac{\cos[2\pi M(l-l')]}{|l-l'|^{2K}}.
\end{eqnarray}
Note that they can be obtained from Eqs.~(\ref{eq:CxLchn}) and
(\ref{eq:CzLchn}) by replacing $q$ and $M_{\rm ch}$ with
$2\pi(M-1/2)$ and $M$, respectively.
On the other hand, the correlations
$\langle S_{0,l}^x S_{0,l'}^x \rangle$
and $\langle S_{\pi,l}^z S_{\pi,l'}^z \rangle$ decay exponentially
because $S^x_{0,l}$ and $S^z_{\pi,l}$ always create
the high-energy rung states
$|t_l^0 \rangle$ and $|t_l^- \rangle$
as a virtual excited state.

Next, we consider the opposite case, the weak-coupling limit ($j\ll1$).
The system in this limit has been investigated with the Abelian
bosonization method.\cite{Chi-Gia,Gia-Tsv,Furu-Zhn}
In these studies, two chains are first bosonized independently,
and then the interchain coupling $J_\perp$ is treated
perturbatively.\cite{Shel}
Four bosonic fields $\phi_\pm(x)$ and $\tilde{\phi}_\pm(x)$ are
introduced, where $\phi_+$ and $\tilde{\phi}_+$
($\phi_-$ and $\tilde{\phi}_-$) are the symmetric (antisymmetric)
combinations of bosonic fields of each chain.
All the fields are massive\cite{Shel} when $h < h_{c1}$.
In the IC regime of $h_{c1} \le h \le h_{c2}$, on the other hand,
the fields $\phi_+$ and $\tilde{{\phi}}_+$ become massless
while the fields $\phi_-$ and $\tilde{{\phi}}_-$ remain massive.
The low-energy effective Hamiltonian for the gapless modes
has the same form
as that of the $S=1/2$ $XXZ$ chain, Eq.~(\ref{eq:HchnBos}).
Furthermore, the spin correlation functions
$\langle S_{\pi,l}^x S_{\pi,l'}^x \rangle$ and
$\langle S_{0,l}^z S_{0,l'}^z \rangle$ have the same $r$ dependence
as in the strong-coupling limit
(but with different values of $K$, $a$, $b$, and
$c$).\cite{Furu-Zhn,note}
The correlators
$\langle S_{0,l}^x S_{0,l'}^x \rangle$ and
$\langle S_{\pi,l}^z S_{\pi,l'}^z \rangle$
decay exponentially,\cite{Furu-Zhn} because they involve the massive
fields.
This result also matches the strong-coupling limit.
Moreover, the incommensurate wavenumber for the short-ranged
correlators is $\tilde q=\pi M$, which is different from the IC
wavenumber for the quasi-long-ranged correlators $q=2\pi(M-1/2)$.
For example, it was found that\cite{Furu-Zhn}
\begin{equation}
\langle S_{\pi,l}^z S_{\pi,l'}^z\rangle = \widetilde{A}_z
(-1)^{l-l'}e^{-|l-l'|/\xi}\frac{\cos[\pi M(l-l')]}{|l-l'|^{1/2+1/4\eta}},
\label{SpiSpi}
\end{equation}
where $\xi$ is a correlation length for a massive mode and
$\widetilde{A}_z$ is a constant.

We have seen that, both in the strong- and weak-coupling limits, the
low-energy physics of the two-leg ladder in the gapless regime is in
the same universality class as the $XXZ$ chain in a magnetic field.
In particular, the spin correlation functions
$\langle S^x_{\pi,l}S^x_{\pi,l'}\rangle$ and
$\langle S^z_{0,l}S^z_{0,l'}\rangle$ and the local magnetization
$\langle S^z_{0,l}\rangle$ have the same forms as the
corresponding functions in the $XXZ$ chain, but with the shifted
wavenumber $2\pi(M-1/2)$ and with different values of
$K$, $a$, $b$, and $c$.\cite{nonuniversal}
It is then very natural to postulate that the universality is not
restricted to the two limits but holds for any $j$.
This allows us to use Eqs.~(\ref{eq:Cxlad})--(\ref{eq:Szlad}) for
analyzing the correlation functions in the ladder for any $j$ and
$M$ ($0<M<1$).
We can thus determine the TL-liquid parameter $K$ of the ladder
by fitting the numerical data to
Eqs.~(\ref{eq:Cxlad})--(\ref{eq:Szlad}) in the same way as we did for
the $XXZ$ chain.
The result is presented in the next subsection.
Finally, we may expect that the $|l-l'|$ dependence of the short-ranged
correlators
$\langle S_{0,l}^x S_{0,l'}^x\rangle$ and
$\langle S_{\pi,l}^z S_{\pi,l'}^z\rangle$ obtained in the weak-coupling
analysis, such as Eq.\ (\ref{SpiSpi}), should also be valid for any $j$
and $M$ ($0<M<1$).

\subsection{Numerical Results}
Here we present the result of the DMRG calculation of the two-spin
correlations
$\langle S_{\pi,l}^x S_{\pi,l'}^x \rangle$ and
$\langle S_{0,l}^z S_{0,l'}^z \rangle$ and the local magnetization
$\langle S_{0,l}^z\rangle$ in the ladder.
Taking $R$, $a$, $b$, and $c$ as free parameters, we fit the data to
the formulas (\ref{eq:Cxlad})--(\ref{eq:Szlad}) for several values of
$j$ and estimate the $M$ dependence of $K = 1/(4 \pi R^2)$.
The numerical calculations were performed for the open ladder
${\cal H} + {\cal H}'$ of $L=100$ rungs ($200$ sites)
for $j = 10.0, 2.0, 1.0, 0.5$ and $\Delta = 1.0, 0.5, 0.0$
using the finite-system DMRG method of the improved version.
We calculated the two-spin correlations for $l = r_0 - r/2$ and
$l' = r_0 + r/2$, where $r_0 = L/2$ for even $r$ and $r_0 = (L+1)/2$
for odd $r$.
The maximum value of the kept states $m$ is $160$.
  From the difference between the data with $m = 160$ and
those with $m = 120$, we estimate the numerical error
due to the truncation.
The estimated errors for $\langle S^x_{\pi,l} S^x_{\pi,l'} \rangle$,
$\langle S^z_{0,l} S^z_{0,l'} \rangle$,
and $\langle S^z_l \rangle$ are, at largest, of order
$10^{-4}$, $10^{-6}$, and $10^{-6}$,
which is almost negligible.
In the course of the calculation, we optimized the value of the extra
boundary field $h'$ to minimize
the effect of the boundary field.\cite{addH}
For finite $j$, however, the boundary effect cannot be eliminated
completely since it can be represented
by the form of ${\cal H}'$ only in the strong-coupling limit $j\gg1$.
As a result, the two-spin correlation functions and the local
magnetization in the open ladder ${\cal H} + {\cal H}'$ might deviate
from the expected form, Eqs.~(\ref{eq:Cxlad})--(\ref{eq:Szlad}),
near the boundaries.
For this reason we used data of smaller range of $r$ for the
fitting than in Sec.~II to reduce the unwanted boundary effect.
We chose the regions
$10 \le r \le 70$, $10 \le r \le 80$, $20 \le r \le 70$, and
$20 \le r \le 80$ for the fitting of the two-spin correlators
and $10 \le l \le 90$ and $20 \le l \le 80$ for the local magnetization.
As in Sec.~II, we regard the mean and
the variance of the fitting parameters
obtained for these different ranges of $r$ and $l$
as the estimated value and the error of the estimates,
respectively.
Incidentally, we have also checked for $(j,M)=(10.0, 0.5)$ that,
without the boundary field $h'$, the local magnetization
$\langle S^z_{0,l}\rangle$ has the Friedel oscillations induced by the
effective boundary field.
We found that the oscillations decay algebraically into the bulk with
the exponent $K$,
as expected from the bosonization analysis.\cite{Affleck}

\begin{figure}
\epsfxsize=80mm
\epsfbox{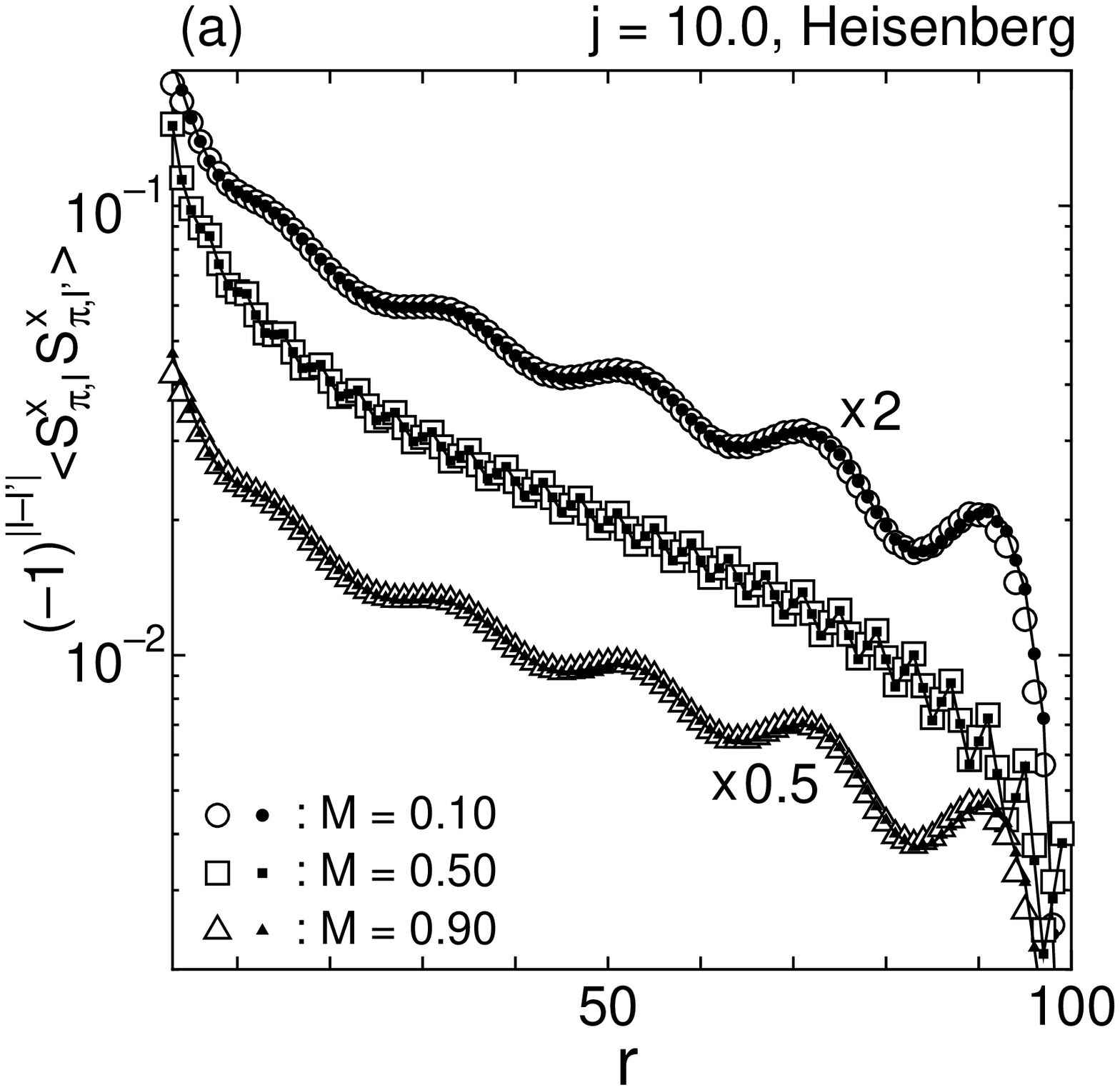}
\epsfxsize=80mm
\epsfbox{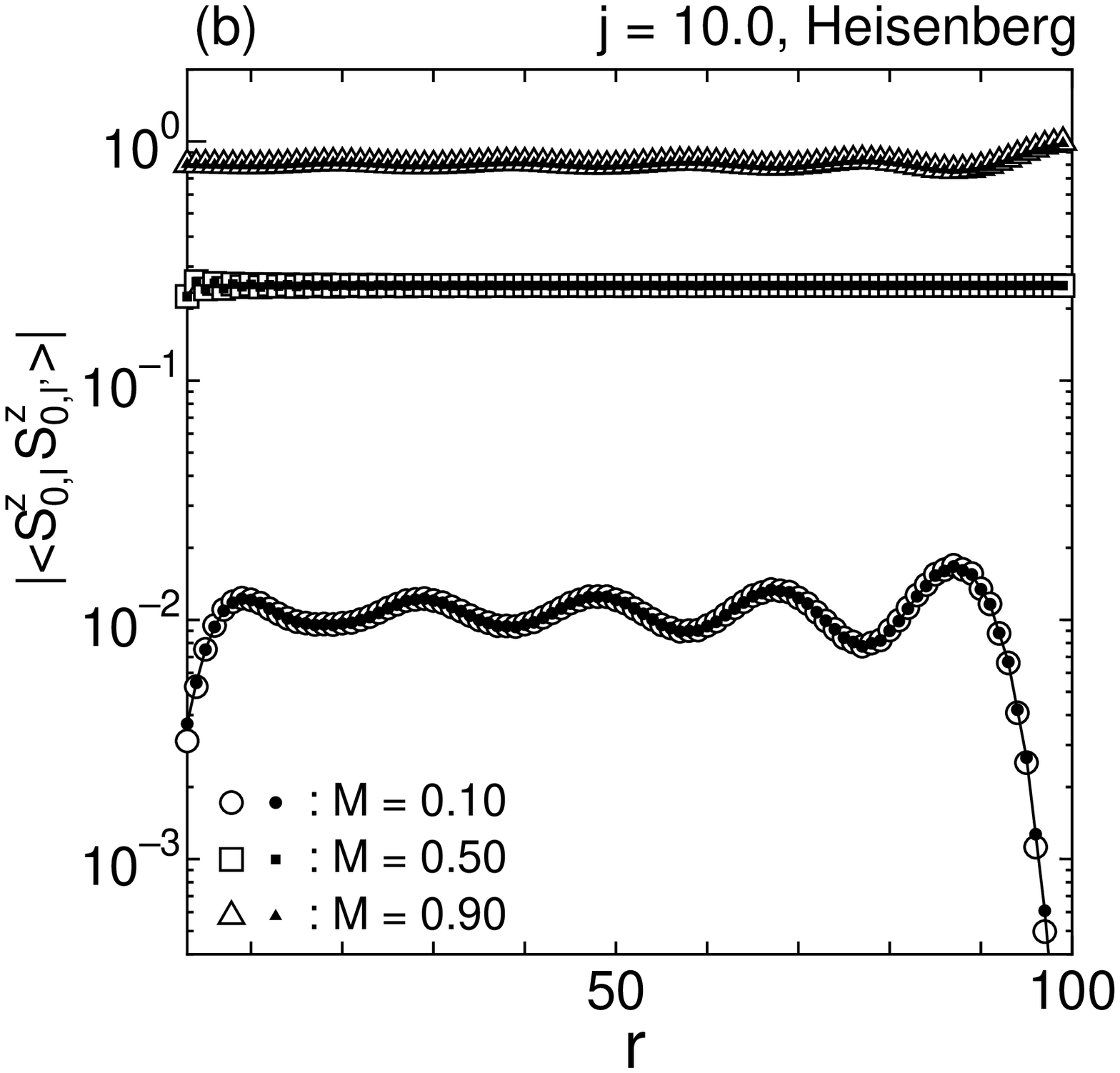}
\epsfxsize=80mm
\epsfbox{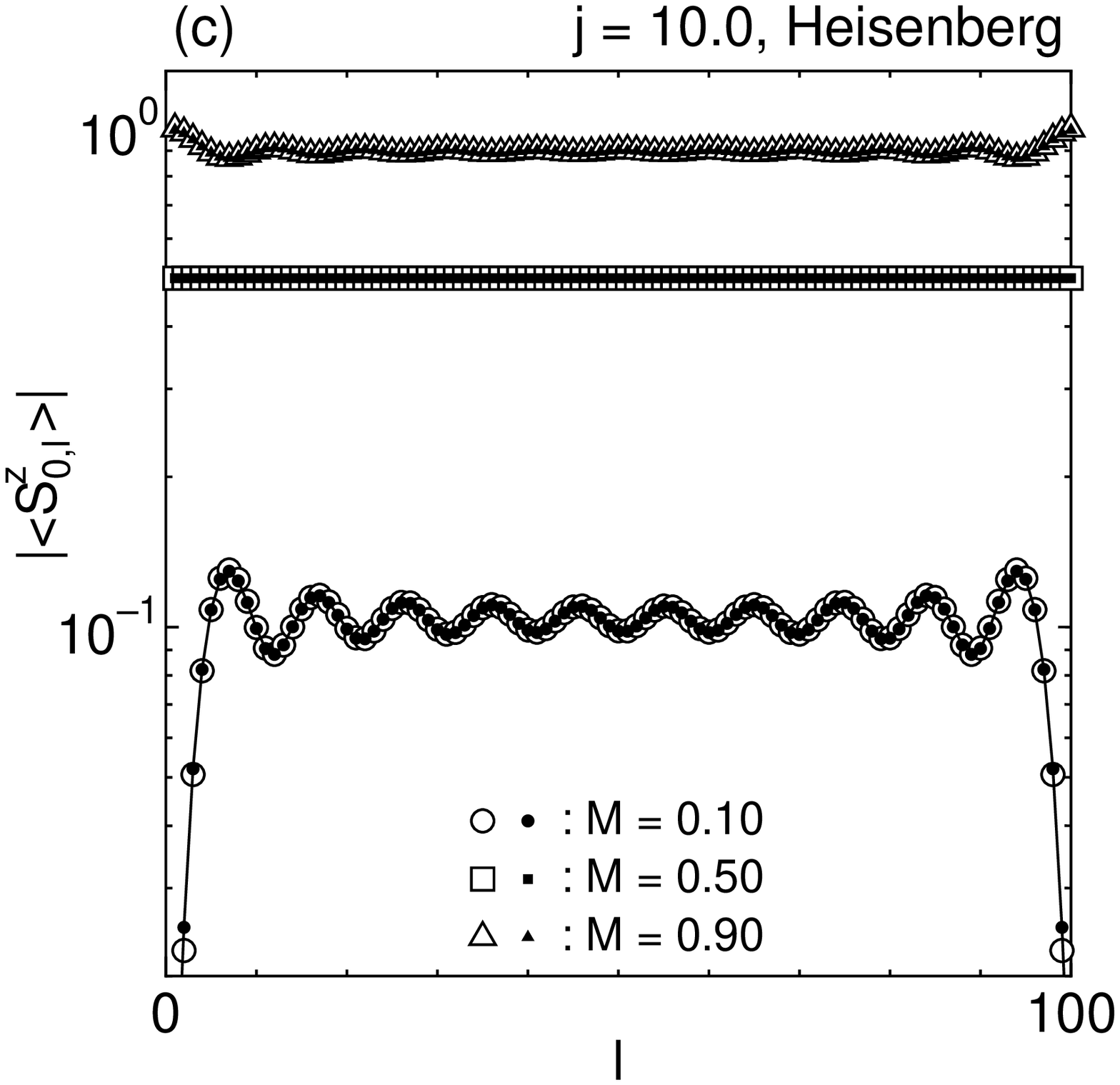}
\caption{(a) $(-1)^{|l-l'|} \langle S_{\pi,l}^x S_{\pi,l'}^x \rangle$
versus $r = |l-l'|$,
(b) $|\langle S_{0,l}^z S_{0,l'}^z \rangle|$ versus $r$,
(c) $\langle S_{0,l}^z \rangle$ versus $l$ for
$j = 10.0$ and the Heisenberg case ($\Delta = 1.0$).
The open symbols are the DMRG data and small solid symbols are
the fits.
The numerical errors of the DMRG data are smaller than the size of
the open symbols.
The data of $(-1)^{|l-l'|} \langle S_{\pi,l}^x S_{\pi,l'}^x \rangle$
for $M=0.10$ and $0.90$ in figure (a)
are multiplied by a factor of $2$ and $0.5$, respectively.}
\label{fig:lad100}
\end{figure}

\begin{figure}
\epsfxsize=80mm
\epsfbox{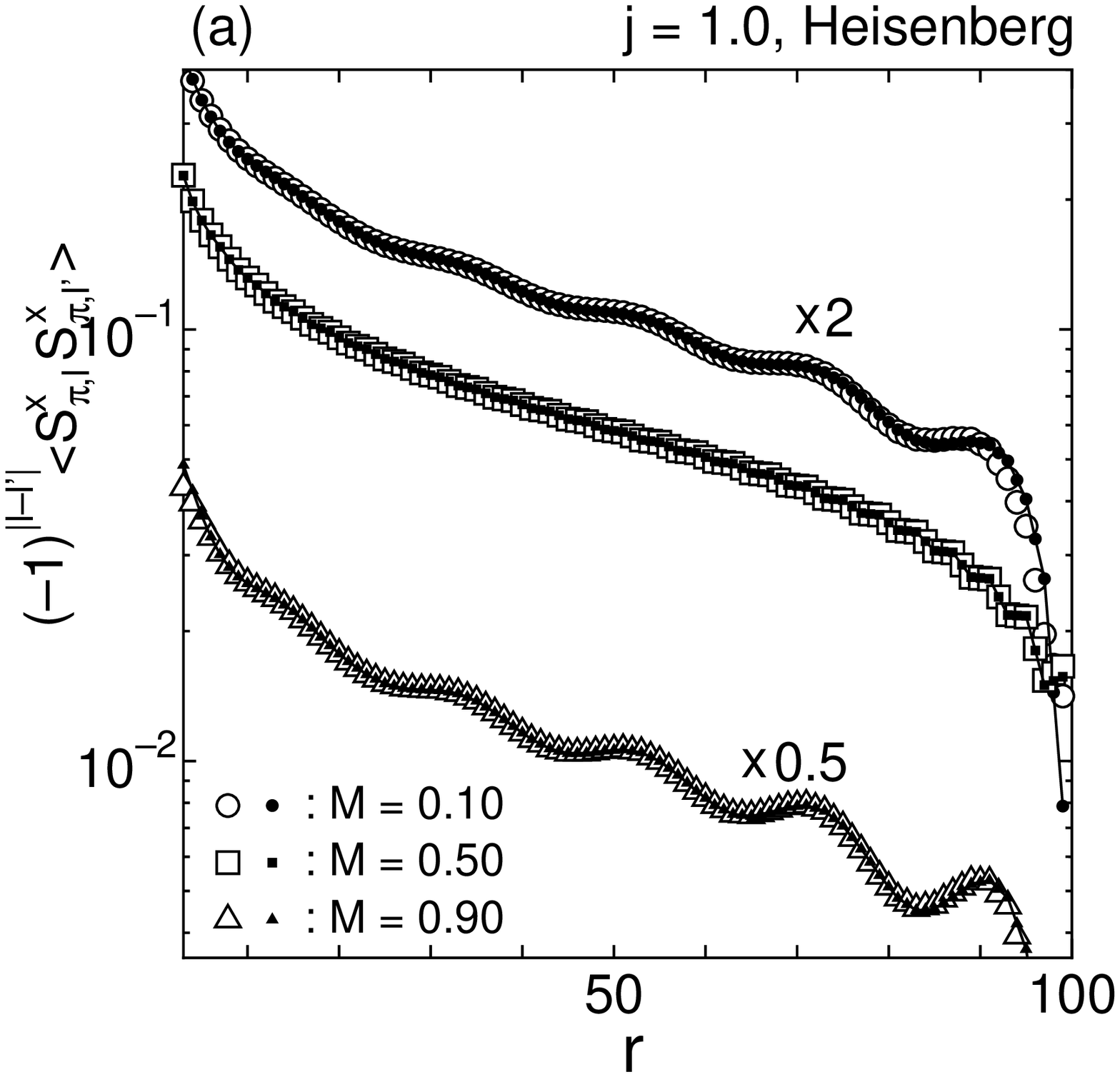}
\epsfxsize=80mm
\epsfbox{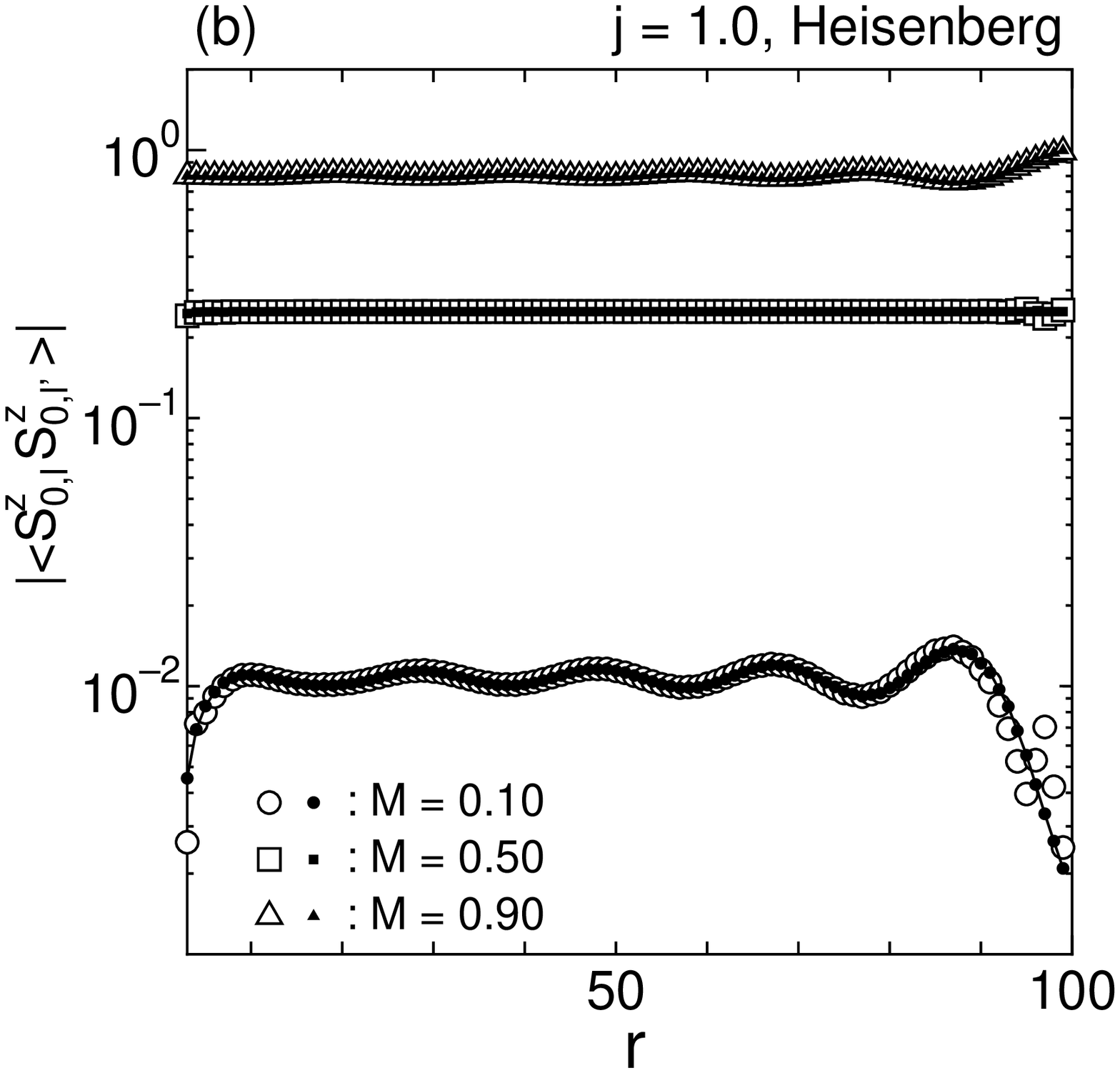}
\epsfxsize=80mm
\epsfbox{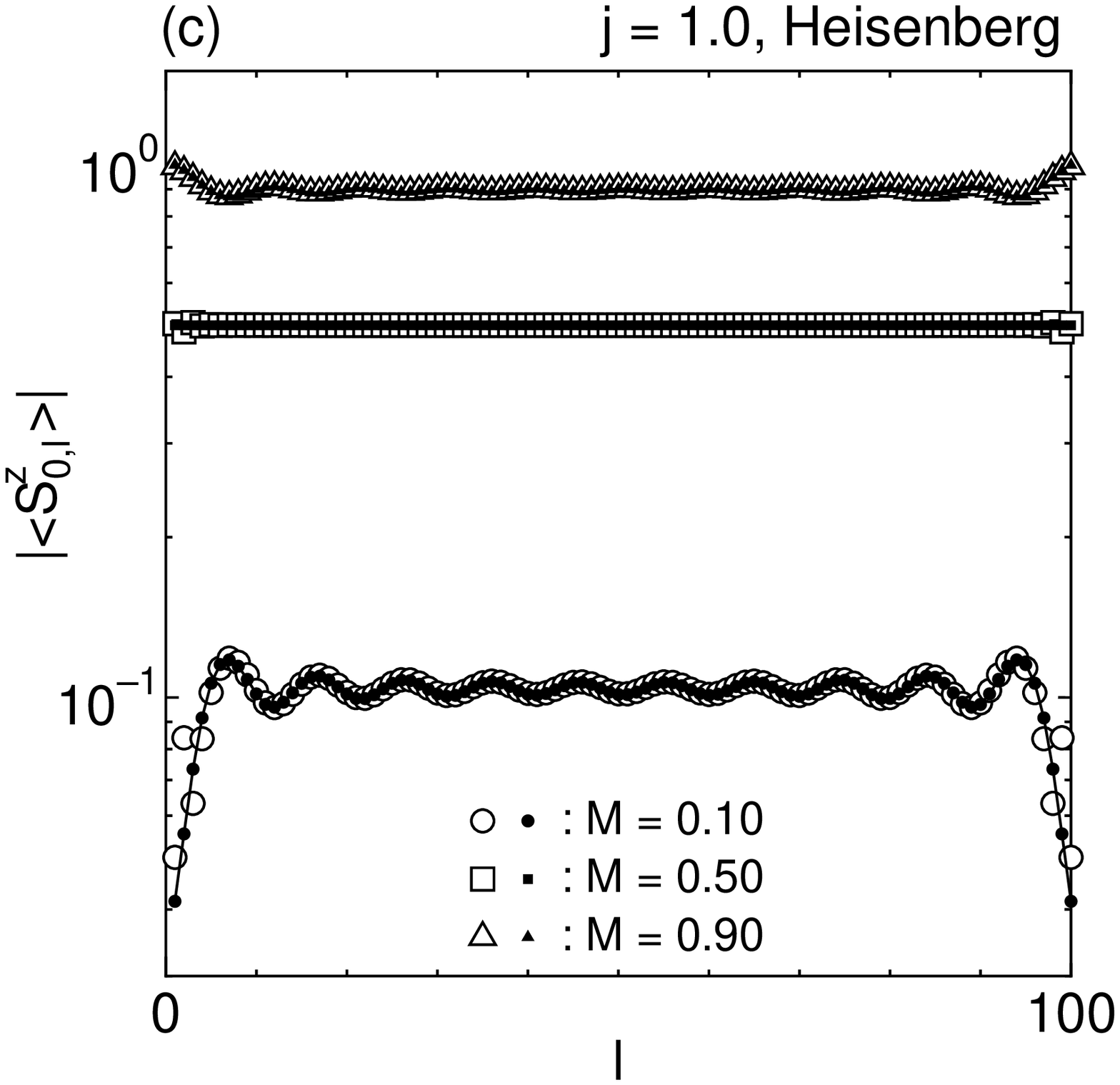}
\caption{(a) $(-1)^{|l-l'|} \langle S_{\pi,l}^x S_{\pi,l'}^x \rangle$
versus $r = |l-l'|$,
(b) $|\langle S_{0,l}^z S_{0,l'}^z \rangle|$ versus $r$,
(c) $\langle S_{0,l}^z \rangle$ versus $l$ for
$j = 1.0$ and the Heisenberg case ($\Delta = 1.0$).
The open symbols are the DMRG data and small solid symbols are
the fitting results.
The numerical errors of the DMRG data are smaller than the size of
the open symbols.
The data of $(-1)^{|l-l'|} \langle S_{\pi,l}^x S_{\pi,l'}^x \rangle$
for $M=0.10$ and $0.90$ in figure (a)
are multiplied by a factor of $2$ and $0.5$, respectively.}
\label{fig:lad010}
\end{figure}

The numerical data of $\langle S_{\pi,l}^x S_{\pi,l'}^x \rangle$,
$\langle S_{0,l}^z S_{0,l'}^z \rangle$, and $\langle S_{0,l}^z\rangle$
for $j = 10.0$ and $1.0$ with $\Delta = 1.0$ (Heisenberg case)
are shown in Figs.~\ref{fig:lad100} and \ref{fig:lad010}
by open symbols, whose sizes are larger than the truncation error
mentioned above.
The small solid symbols in the figures are the fits
to the DMRG data of the two-spin correlations for $10 \le r \le 80$
and to those of the local magnetization for $10 \le l \le 90$, respectively.
It is clearly seen that the fitting works extremely well
for $j = 10.0$, confirming the validity of the formulas
(\ref{eq:Cxlad}), (\ref{eq:Czlad}), and (\ref{eq:Szlad}).
Furthermore, the agreement between the numerical data and the fits
at $j = 1.0$ is also quite good except some deviations near the
boundary, indicating that the formulas are accurate for the
intermediate-coupling regime of $j$ as well.
We note that the quality of the fitting is also good for other values
of $j$ and $\Delta$ that we have examined.
We therefore conclude that the gapless mode of the two-leg ladders
is a TL liquid for arbitrary $j$,
and accordingly, the properties of the strong- and weak-coupling
ladders are smoothly connected.

\begin{figure}
\epsfxsize=80mm
\epsfbox{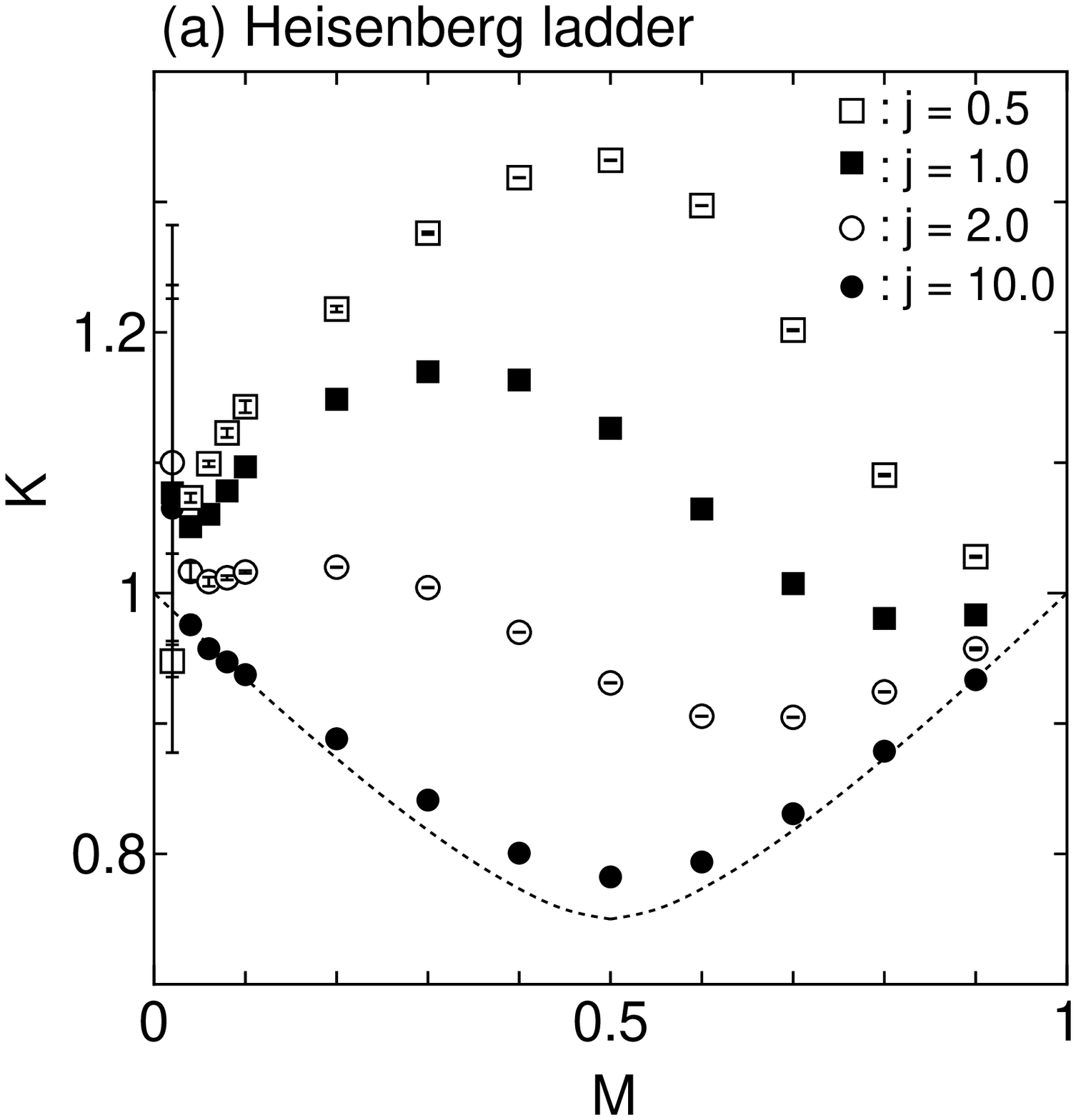}
\epsfxsize=80mm
\epsfbox{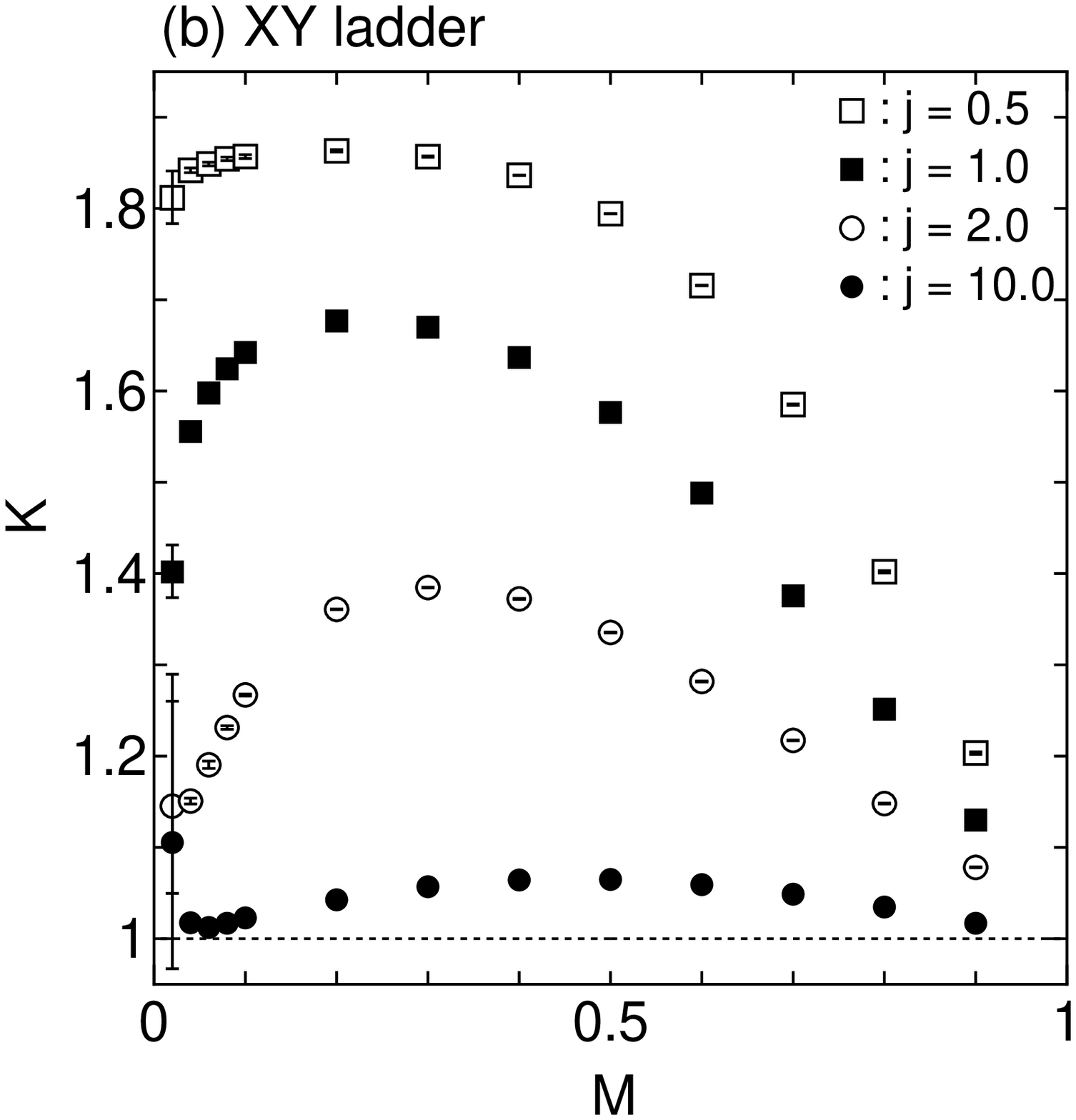}
\caption{The TL-liquid parameter $K = 1/(4 \pi R^2)$ estimated from
the fitting of $\langle S^x_{\pi,l} S^x_{\pi,l'} \rangle$ for
$j = 10.0, 2.0, 1.0, 0.5$ and
for (a) the Heisenberg ladder ($\Delta = 1.0$) and
(b) the $XY$ ladder ($\Delta = 0$).
The dotted curves represent the exact values of the $S=1/2$ $XXZ$
chain with the anisotropy $\Delta/2$.}
\label{fig:Klad}
\end{figure}

Next we show in Fig.~\ref{fig:Klad} the $M$ dependence of
$K$ estimated from the data of
$\langle S_{\pi,l}^x S_{\pi,l'}^x \rangle$
for various $j$ in both the Heisenberg ($\Delta = 1.0$) and
$XY$ ($\Delta = 0$) cases.
In the earlier study\cite{Usami} the exponent $\eta$ was
obtained for $j=5.0$ in the Heisenberg case only.
We note that the estimation from
$\langle S_{\pi,l}^x S_{\pi,l'}^x \rangle$
is more reliable than that from
$\langle S_{0,l}^z S_{0,l'}^z \rangle$
or $\langle S_{0,l}^z\rangle$ as we have seen in the $XXZ$ chain.
Theoretically\cite{Furu-Zhn} it is expected that $K$ should
approach the universal value $K=1$ when $M\to0$ as well as when
$M\to1$, since the system is equivalent to the dilute limit of
hard-core bosons.
Although the data for $M\to0$ have large error bars,
we may conclude that our results for the Heisenberg case are
consistent with the theoretical prediction.
Our results for the $XY$ case show a more subtle feature.
At first sight the results for weaker couplings ($j=1.0$ and 0.5) do
not seem to approach $K=1$ as $M\to0$.
We think, however, that $K$ changes very rapidly at small $M$ to
approach $K=1$, in view of the data for $j=10.0$ and $2.0$, which
are consistent with the theory.
Unfortunately, it is difficult to numerically estimate $K$ for small
$M$ with high accuracy to resolve this issue.

In the strong-coupling limit $j\gg1$,
the ladder system with anisotropy $\Delta$ is
equivalent to the $S=1/2$ $XXZ$ chain with anisotropy $\Delta/2$,
as explained in the previous subsection.
Figure \ref{fig:Klad} clearly shows that
the estimated value of $K$ for $j = 10.0$ is consistent with
the anticipated behavior shown
as the dotted curves in both the Heisenberg and $XY$ cases.
As $j$ decreases, $K$ increases monotonically for any $M$ ($0<M<1$).
Thus, $K$ is always larger than 1 in the $XY$ ladder
because $K\to1$ for any $M$ in the large $j$ limit.
In the Heisenberg ladder, on the other hand, $K$ is smaller than 1 in
the strong-coupling limit, as expected from the mapping to the $XXZ$
chain.
Upon decreasing the interchain coupling $j$, $K$ starts to increase
and the $K$-$M$ relation changes from a concave curve to a convex one.
We note that the similar behavior is observed
also in the ladder with $\Delta = 0.5$: As $j$ decreases,
the $K$-$M$ relation changes from a concave curve at $j\gg1$,
corresponding to the behavior of the $XXZ$ chain
with anisotropy $\Delta/2 = 0.25$, to a convex one.
We thus consider that this behavior of the $K$-$M$ curve
is an universal feature for $0 < \Delta \le 1$.

The TL-liquid parameter $K$ determines the long-distance behavior of
correlation functions in the thermodynamic limit.
For example, the leading term of the correlator
$\langle S^z_{0,l} S^z_{0,l'}\rangle - M^2$
decays as $\cos[2\pi M(l-l')]/|l-l'|^{2K}$ for $K<1$,
while it decays like $|l-l'|^{-2}$ for $K>1$.
Hence, in the ladder with $0<\Delta\le1$
the leading term of the correlator changes
from $\cos[2\pi M(l-l')]/|l-l'|^{2K}$ to $|l-l'|^{-2}$
at a critical value $j_c(M)$ as $j$ decreases,
while in the $XY$ ladder the leading term
is always $|l-l'|^{-2}$.
On the other hand, the correlator
$\langle S^x_{\pi,l} S^x_{\pi,l'}\rangle$ decays as
$(-1)^{l-l'}/|l-l'|^{1/2K}$ in the whole range of $K$ covered
in Figs.~\ref{fig:Klad} (a) and \ref{fig:Klad} (b).
The temperature dependence of the spin-lattice relaxation rate $1/T_1$
in NMR experiments is directly related to the TL-liquid parameter $K$
through the decay exponent of the most slowly decaying
correlation.\cite{Chi-Gia}
  From the behavior of $K$-$M$ relation obtained above,
we find that the correlator $\langle S^x_{\pi,l} S^x_{\pi,l'}\rangle$
decays most slowly for any $j$, $M$, and $0 \le \Delta \le 1$.
We therefore conclude that at low temperatures the relaxation rate
of the ladder in the gapless regime always shows a power-law divergence
$1/T_1\propto T^{-1+(1/2K)}$.

\begin{figure}
\epsfxsize=80mm
\epsfbox{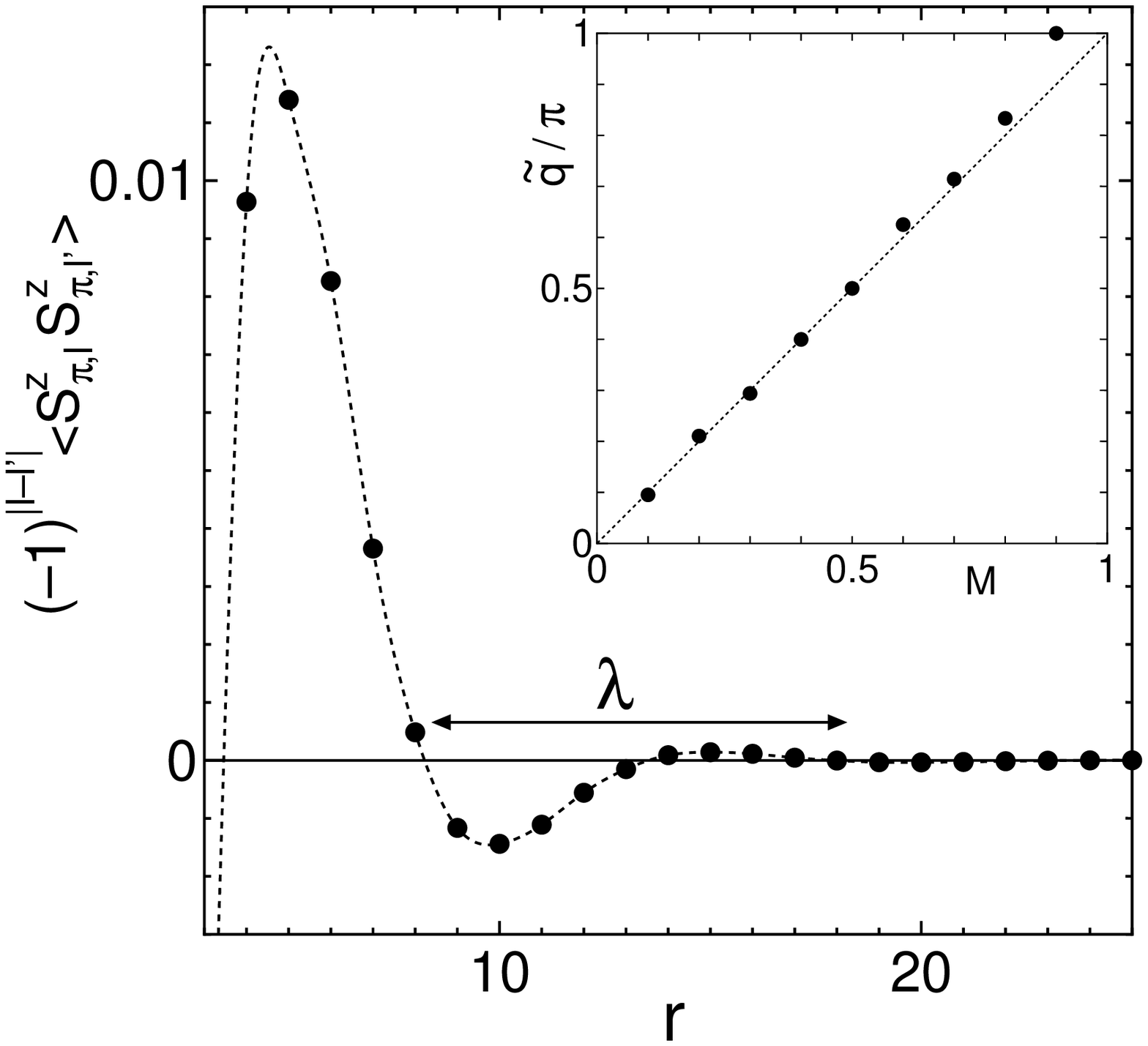}
\caption{The correlation function
$(-1)^{|l-l'|} \langle S^z_{\pi,l}S^z_{\pi,l'}\rangle$ 
in the Heisenberg ladder at $j=0.5$ and $M=0.2$.
The dotted curve is a guide to the eye.
Inset: The $M$ dependence of the IC wavenumber $\tilde{q}$
of $\langle S^z_{\pi,l}S^z_{\pi,l'}\rangle$ in the Heisenberg ladder
with $j = 0.5$.
The dotted line represents the theoretical prediction
$\tilde{q}/\pi = M$.
}
\label{fig:SzSz}
\end{figure}

Figure \ref{fig:SzSz} shows the numerical result of
$\langle S^z_{\pi,l}S^z_{\pi,l'}\rangle$ for the Heisenberg ladder
at $j=0.5$.
It exhibits exponentially decaying oscillatory behavior.
 From the period $\lambda$ of oscillations,
we obtain the IC wavenumber $\tilde q = 2\pi/\lambda$
as a function of $M$; see the inset figure.
The result confirms the theoretical prediction
$\tilde q=\pi M$.
This IC wavenumber $\tilde q$ tells us that the massive magnon
dispersion has a minimum excitation energy at\cite{Furu-Zhn}
$q=\pi-\tilde q=\pi(1-M)$.
Accordingly, the dynamical spin structure factor
$S^{zz}_\pi(q,\omega)$ should have a power-law divergence along the
energy dispersion which is roughly shifted by $\pi M$ from that of
the triplet magnon dispersion in the absence of the magnetic field.
It would be interesting if this feature is observed by inelastic
neutron scattering experiments.

\section{CONCLUSIONS}
In this paper we have studied the ground-state spin correlations
in the gapless IC regime of the $S=1/2$ $XXZ$ chain and the two-leg
AF ladder in a magnetic field.
We have used the $S=1/2$ $XXZ$ chain as a first test ground to apply
the method we developed in our previous work:
We numerically computed the two-spin correlation functions and the
local magnetization by the DMRG method and fit the results to
functions which are obtained using the bosonization technique.
The fitting parameters are the TL-liquid parameter $K$ and the
amplitudes of bosonic operators.
We found good agreement between $K$ estimated from the fitting and
$K$ calculated from the Bethe ansatz.
As a byproduct we obtained the amplitudes of the dominant terms in
$\langle S^x_l S^x_{l'}\rangle$ and $\langle S^z_l S^z_{l'}\rangle$.

We have applied the same technique to
the two-leg AF ladder in the gapless IC regime.
It has been known that in both the strong- and weak-coupling limits
the low-energy excitations in the ladder are regarded
as a TL liquid like the $XXZ$ chain in a field.
We fit our DMRG data of the two-spin correlation functions and the
local magnetization of the ladder to the same bosonization
formulas we used in the analysis of the $XXZ$ chain.
The fitting worked very well not only in the strong- and weak-coupling
limits but for broad range of the interchain coupling strength $j$.
We thereby confirmed that the low-energy gapless excitations are
indeed described as the TL liquid for any $j$ and
the properties of the strong- and weak-coupling ladders
are smoothly connected.
For several values of $j$, we have determined $K$,
which shows nontrivial $j$ and $M$ dependences
(Fig.~\ref{fig:Klad}).
It turned out that, for any $M$ ($0<M<1$),
$K$ is a monotonically decreasing function of $j$.
In the ladder with anisotropy $0 < \Delta \le 1$
the $K$-$M$ relation changes from a concave curve at $j\gg1$
to a convex one as $j$ decreases,
while in the $XY$ ladder ($\Delta=0$)
it changes from a line $K=1$ at $j\gg1$ to a convex curve.
We also found that the spin-lattice relaxation rate
in NMR measurement shows a power-law divergence
$1/T_1 \propto T^{-1+(1/2K)}$ at low temperature for any $j$.

\acknowledgements

Numerical computations were performed at the Yukawa Institute
Computing Facility.
The work of AF was in part supported by Grant-in-Aid for Scientific
Research on Priority Areas (A) from the Ministry of Education,
Science, Sports and Culture (No.~12046238) and by Grant-in-Aid for
Scientific Research from the Japan Society for the Promotion of
Science (No.~11740199).

\appendix
\section*{Derivation of Correlators}

We briefly explain the derivation of Eqs.~(\ref{eq:Cxechn}) and
(\ref{eq:Czechn}).
As is always the case  with the bosonization, we need to introduce a
short-distance cutoff to obtain finite results.
The lattice spacing in the original Hamiltonian serves as
the natural cutoff scale.

When we use Eqs.~(\ref{eq:Szchn}) and (\ref{eq:S-chn}) with the mode
expansions (\ref{eq:mode1}) and (\ref{eq:mode2}), we encounter the
summation
\begin{equation}
\sum^\infty_{n=1}\frac{1}{n}
   \left[1-\cos\left(\frac{\pi nl}{L+1}\right)\right],
\end{equation}
which is formally divergent.
We regularize it by inserting an exponential factor
$e^{-\pi n/(L+1)}$:
\begin{equation}
\sum^\infty_{n=1}\frac{1}{n}e^{-\pi n/(L+1)}
\left[1-\cos\left(\frac{\pi nl}{L+1}\right)\right]
=\ln[f(l)],
\label{ln(f)}
\end{equation}
where $f(l)\equiv f_1(l)$ defined in Eq.~(\ref{eq:fx}).
We note that Eq.~(\ref{ln(f)}) is a very good approximation except
near the points where $f_1(l)$ is divergent.
Taking derivatives with respect to $l$, we obtain
\begin{equation}
\sum^\infty_{n=1}e^{-\pi n/(L+1)}\sin\left(\frac{\pi nl}{L+1}\right)
=\frac{1}{2}\cot\left(\frac{\pi l}{2(L+1)}\right)
\end{equation}
and
\begin{equation}
\sum^\infty_{n=1}n e^{-\pi n/(L+1)}\cos\left(\frac{\pi nl}{L+1}\right)
=-\frac{1}{4\sin^2\left(\frac{\pi l}{2(L+1)}\right)}.
\end{equation}
Another point to note is that for $\varepsilon_i=\pm1$
\begin{eqnarray}
\langle e^{i2\pi R\epsilon_1\tilde\phi(l)}
   e^{i2\pi R\epsilon_2\tilde\phi(l')}\rangle
&\propto&
   \int^{1/R}_0 R e^{i2\pi R(\epsilon_1+\epsilon_2)\tilde\phi_0}
    d\tilde\phi_0 \nonumber\\
&\propto&
\delta_{\epsilon_1+\epsilon_2,0}.
\end{eqnarray}

With the above-mentioned formulas, it is straightforward to obtain
\begin{eqnarray*}
&&
\langle\cos[2\pi R\tilde\phi(l)]\cos[2\pi R\tilde\phi(l')]\rangle
=\frac{[f(2l)f(2l')]^{\eta/2}}{2[f(l-l')f(l+l')]^\eta},
   \\
&&
\langle e^{i2\pi R\epsilon_1\tilde\phi(l)}
   e^{-i2\pi R\epsilon_1\tilde\phi(l')}e^{i\epsilon_2\phi(l')/R}\rangle
\\
&&
=i\epsilon_1\epsilon_2 e^{iq\epsilon_2 l'}
\frac{{\rm sgn}(l-l')[f(2l)f(2l')]^{\eta/2}}
       {[f(l-l')f(l+l')]^\eta [f(2l')]^{1/2\eta}},
\\
&&
\langle e^{i2\pi R\epsilon_0\tilde\phi(l)}e^{i\epsilon_1\phi(l)/R}
   e^{-i2\pi R\epsilon_0\tilde\phi(l')}e^{i\epsilon_2\phi(l')/R}\rangle
\\
&&
=-\epsilon_1\epsilon_2 e^{iq(\epsilon_1l+\epsilon_2l')}
\frac{[f(2l)f(2l')]^{\eta/2-1/2\eta}}
       {[f(l-l')f(l+l')]^\eta}
\left(\frac{f(l-l')}{f(l+l')}\right)^{\epsilon_1\epsilon_2/\eta},
\\
&&
\left\langle\frac{d\phi}{dl}\frac{d\phi}{dl'}\right\rangle
=-\frac{1}{2\pi}\left(\frac{1}{f_2(l-l')}+\frac{1}{f_2(l+l')}\right)
+\left(\frac{q}{2\pi}\right)^2,
\\
&&
\left\langle\left(\frac{d\phi}{dl}-\frac{q}{2\pi}\right)
              \sin\frac{\phi(l')}{R}\right\rangle
=\frac{\cos(ql')}{2\pi R}\frac{g(l+l')-g(l-l')}{[f(2l')]^{1/2\eta}},
\\
&&
\langle e^{i\epsilon_1\phi(l)/R} e^{i\epsilon_2\phi(l')/R}\rangle
=\frac{e^{iq(\epsilon_1l+\epsilon_2l')}}{[f(2l)f(2l')]^{1/2\eta}}
   \left(\frac{f(l-l')}{f(l+l')}\right)^{\epsilon_1\epsilon_2/\eta}.
   \\
\end{eqnarray*}

\end{multicols}

\begin{table}
\caption{
The correlation amplitudes;
(a) $A_x = c^2/2$ estimated from the data of
$\langle S^x_l S^x_{l'} \rangle$;
(b) $A_z = a^2/2$ estimated from the data of
$\langle S^z_l \rangle$.
The figures in parentheses indicate the error bar
on the last quoted digits.
The error bars of $A_z$ for $\Delta=0$ and $0.05 \le M \le 0.45$
are smaller than $10^{-5}$.
The exact values for $M_{\rm ch} = 0$ given by
Eqs. (\ref{eq:LZx}) and (\ref{eq:LZz}) are also listed.
}
\label{tab:Achn}
(a)$A_x$
\begin{tabular}{ccccccccccc}
$M_{\rm ch}$    &    0      &    0.05   &    0.10   &    0.15   &    0.20
   &    0.25   &    0.30   &    0.35   &    0.40   &    0.45   \\ \hline
$\Delta= 0.0$ & 0.14709   & 0.14626(1)& 0.14364(6)& 0.1390(1) & 0.13262(1)
   & 0.12410(6)& 0.1132(3) & 0.0993(7) & 0.081(2)  & 0.0594(7) \\
$\Delta= 0.1$ & 0.14451   & 0.14369(7)& 0.1413(1) & 0.1371(1) & 0.13101(7)
   & 0.12293(2)& 0.1125(3) & 0.0991(7) & 0.081(2)  & 0.0597(7) \\
$\Delta= 0.2$ & 0.14187   & 0.1408(4) & 0.1390(2) & 0.1351(2) & 0.1294(1)
   & 0.12174(5)& 0.1111(7) & 0.0988(6) & 0.081(2)  & 0.0600(7) \\
$\Delta= 0.3$ & 0.13921   & 0.1384(3) & 0.1366(3) & 0.1330(3) & 0.1278(2)
   & 0.12053(9)& 0.1111(2) & 0.0985(6) & 0.081(2)  & 0.0601(7) \\
$\Delta= 0.4$ & 0.13656   & 0.1358(3) & 0.1342(4) & 0.1310(3) & 0.1261(3)
   & 0.1193(1) & 0.1104(1) & 0.0982(6) & 0.081(2)  & 0.0603(7) \\
$\Delta= 0.5$ & 0.13400   & 0.1332(4) & 0.1318(5) & 0.1289(4) & 0.1245(3)
   & 0.1182(2) & 0.10973(9)& 0.0979(5) & 0.081(2)  & 0.0605(7) \\
$\Delta= 0.6$ & 0.13164   & 0.1310(5) & 0.1294(6) & 0.1268(5) & 0.1229(4)
   & 0.1170(2) & 0.10905(7)& 0.0976(5) & 0.081(2)  & 0.0606(7) \\
$\Delta= 0.7$ & 0.12973   & 0.1281(6) & 0.1270(7) & 0.1248(5) & 0.1213(4)
   & 0.1159(3) & 0.10839(6)& 0.0973(5) & 0.081(2)  & 0.0607(7) \\
$\Delta= 0.8$ & 0.12896   & 0.1257(8) & 0.1247(8) & 0.1227(6) & 0.1197(5)
   & 0.1148(3) & 0.10775(7)& 0.0970(5) & 0.081(1)  & 0.0609(7) \\
$\Delta= 0.9$ & 0.13214   & 0.1233(9) & 0.1223(9) & 0.1207(7) & 0.1182(6)
   & 0.1137(3) & 0.10714(8)& 0.0967(4) & 0.081(1)  & 0.0610(8) \\
$\Delta= 1.0$ &           & 0.121(1)  & 0.120(1)  & 0.1188(8) & 0.1177(9)
   & 0.1127(4) & 0.1065(1) & 0.0958(6) & 0.081(1)  & 0.0610(8)
\end{tabular}
(b)$A_z$
\begin{tabular}{ccccccccccc}
$M_{\rm ch}$    &    0      &    0.05   &    0.10   &    0.15   &    0.20
     &    0.25   &    0.30   &    0.35   &    0.40   &    0.45   \\ \hline
$\Delta= 0.0$  & 0.05066   & 0.05066   & 0.05066   & 0.05066   & 0.05066
     & 0.05066   & 0.05066   & 0.05066   & 0.05066   & 0.05066   \\
$\Delta= 0.1$  & 0.05929   & 0.0599(7) & 0.0581(3) & 0.0567(6) & 0.0544(1)
     & 0.0537(7) & 0.0513(6) & 0.0516(6) & 0.049(1)  & 0.0510(8) \\
$\Delta= 0.2$  & 0.06891    & 0.071(1)  & 0.0662(6) & 0.063(1)  & 0.0580(2)
     & 0.056(1)  & 0.052(1)  & 0.052(1)  & 0.048(2)  & 0.051(1)  \\
$\Delta= 0.3$  & 0.07978    & 0.083(2)  & 0.0748(7) & 0.069(1)  & 0.0614(4)
     & 0.059(1)  & 0.052(1)  & 0.053(1)  & 0.048(2)  & 0.052(2)  \\
$\Delta= 0.4$  & 0.09231   & 0.097(3)  & 0.0838(6) & 0.075(1)  & 0.0645(6)
     & 0.060(1)  & 0.053(2)  & 0.053(2)  & 0.047(3)  & 0.052(2)  \\
$\Delta= 0.5$  & 0.10713   & 0.113(5)  & 0.093(4)  & 0.080(1)  & 0.0674(8)
     & 0.062(2)  & 0.053(2)  & 0.053(2)  & 0.046(3)  & 0.052(3)  \\
$\Delta= 0.6$  & 0.12539   & 0.132(6)  & 0.10263(5)& 0.0854(9) & 0.070(1)
     & 0.063(2)  & 0.054(2)  & 0.053(2)  & 0.046(3)  & 0.052(3)  \\
$\Delta= 0.7$  & 0.14930   & 0.153(8)  & 0.1121(5) & 0.0903(4) & 0.072(1)
     & 0.065(2)  & 0.054(2)  & 0.053(2)  & 0.045(4)  & 0.052(3)  \\
$\Delta= 0.8$  & 0.18414   & 0.176(10) & 0.121(1)  & 0.09486(6)& 0.074(2)
     & 0.066(2)  & 0.054(2)  & 0.054(2)  & 0.045(4)  & 0.052(4)  \\
$\Delta= 0.9$  & 0.24844   & 0.20(1)   & 0.131(2)  & 0.0990(7) & 0.076(2)
     & 0.067(1)  & 0.054(2)  & 0.054(2)  & 0.047(4)  & 0.052(4)  \\
$\Delta= 1.0$  &           & 0.23(1)   & 0.139(3)  & 0.103(1)  & 0.078(2)
     & 0.067(1)  & 0.054(3)  & 0.054(2)  & 0.044(4)  & 0.052(4)
\end{tabular}
\end{table}

\newpage
\begin{multicols}{2}

\end{multicols}


\begin{references}
\bibitem{no hat Ax}
We are not sure about the accuracy of our estimates for the amplitude
$\widehat{A}_x$ of the subleading term, and thereby we do not show
them here.
\bibitem{Ampnote}
The estimates of $A_z$ for small $M>0$ shown in Table \ref{tab:Achn} (b)
do not seem to approach smoothly the exact value at $M_{\rm ch}=0$
as $M_{\rm ch} \to 0$.
We do not know exactly the reason why this happens.
This might be due to  the same reason as the one for the deviation
of $K$ estimated from $\langle S^z_l \rangle$ mentioned in the text.
\bibitem{note}
We note that the notation has changed from Ref.\ \onlinecite{Furu-Zhn}.
The parameter $\eta$ in the present paper is equal to the inverse of
the $\eta$ used in Ref.\ \onlinecite{Furu-Zhn}.
\bibitem{nonuniversal}
In the strong-coupling limit, the parameters $K$, $a$, $b$, and $c$
in the ladder with the anisotropy $\Delta$ are
related to those in the $XXZ$ chain with the anisotropy $\Delta/2$;
see Eqs.~(\ref{eq:Cxlad})-(\ref{eq:Qlad}).
\bibitem{addH}
To be more concrete, we optimized the value of $h'$ to
minimize the deviation of
$\langle S^z_{0,l} \rangle$ for $M = 1/2$ from the constant value,
$1/2$; see Eq.~(\ref{eq:Szlad}).
\end{references}
\end{document}